\documentclass[aps,prb,twocolumn,showpacs,floatfix,twoside,superscriptaddress,eqsecnum]{revtex4-2}
\usepackage{amssymb}
\usepackage{graphicx}
\usepackage{amsmath}
\usepackage{amsfonts}
\usepackage{esint}
\usepackage{color}
\usepackage{xcolor}
\usepackage{float}
\usepackage{subcaption}
\usepackage{placeins}
\usepackage[colorlinks=true,citecolor=blue,linkcolor=blue,urlcolor=blue]{hyperref}
\usepackage[justification=centerlast,font={small}]{caption}

\DeclareMathOperator*{\tr}{Tr}

\DeclareMathOperator{\sn}{sn}

\newcommand{\BT}{\mathcal{T}}

\newcommand{\BQ}{\mathcal{Q}}

\newcommand{\BU}{\mathcal{U}}

\newcommand{\BF}{\mathcal{F}}

\begin{document}

\title{Phase transition into Instanton Crystal.}
\date{\today }
\author{Grigory A. Starkov}
\affiliation{Ruhr University Bochum, Faculty of Physics and Astronomy, Bochum, 44780,
Germany}
\email{Grigorii.Starkov@rub.de}
\author{Konstantin B. Efetov}
\affiliation{Ruhr University Bochum, Faculty of Physics and Astronomy, Bochum, 44780,
Germany}
\affiliation{National University of Science and Technology \textquotedblleft
MISiS\textquotedblright, Moscow 119049, Russia}
\email{Konstantin.B.Efetov@rub.de}

\begin{abstract}
We propose a class of models exhibiting instanton crystal phase. In this
phase, the minimum of the free energy corresponds to a configuration with an
imaginary-time-dependent order parameter in a form of a chain of alternating
instantons and antiinstantons. The resulting characteristic feature of this
state is that the average of the order parameter over the imaginary time vanishes.
In order to study the model in a broad
region of parameters of the model quantitatively, and prove the existence of
the instanton crystal phase, we develop an efficient numerical scheme,
suitable for the exact treatment of the proposed models. In a certain limit,
results demonstrating the existence of the instanton crystal phase are obtained also
analytically. The numerical study of the model shows that there is a
phase transition between the instanton crystal and the state with the
imaginary-time-independent order parameter.
\end{abstract}

\maketitle

\section{Introduction.}

The standard way of describing a phase transition is based on the concept of an
order parameter introduced by Landau \cite{landau}. This quantity equals
zero in the disordered phase but is finite in the ordered one. The order
parameter can be scalar, vector, tensor, etc. The beauty of this approach
follows from the universality of the description because the critical
behavior depends on the symmetry of the order parameter rather than on the
details of the interaction.

Although the Landau theory is by construction applicable only near the critical point and not too close to it,
so that the fluctuations can be considered small, the concept of the order parameter provides the means of description of the ordered phase for all the temperatures below the transition temperature.
As such, in the case of $\mathbb{Z}_2$ symmetry breaking, for example, the order parameter is real and the minimum of the free energy corresponds to the two possible values $+1$, $-1$ (if properly rescaled) of the order parameter.


\begin{figure}[t]
 \includegraphics[width = 2.8in]{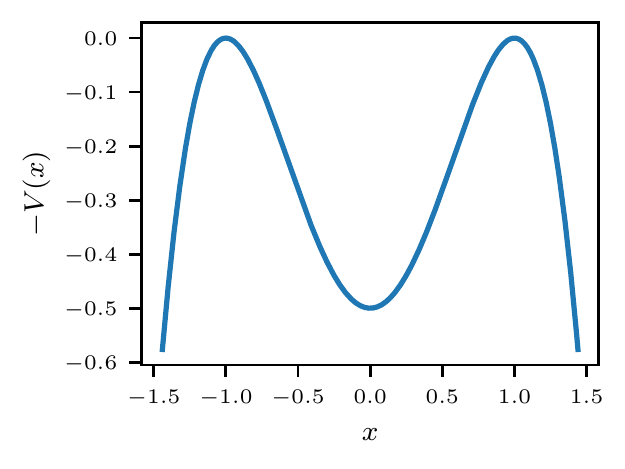}
 \caption{Inverted potential $-V(x) = -\frac12(x^2-1)^2$.}
 \label{inverted_potential}
\end{figure}

The situation becomes more interesting in the quantum limit at low
temperatures, where the tunnelling effects become important.
The toy model which is usually employed to discuss this kind of phenomena is that of a particle moving in a double-well potential.
The convenient way to study the thermodynamics is to use the Euclidean path integral formalism with the imaginary time $\tau$.
The corresponding Euclidean action of the toy model is introduced as
\begin{equation}
 S = \int \mathcal{L} d\tau = \int d\tau \left[\frac12\left(\frac{dx}{d\tau}\right)^2 + \frac12(x^2-1)^2\right],\label{euclidean_action}
\end{equation}
where $x$ is the coordinate of the moving particle.
The lagrangian $\mathcal{L}$ can be interpreted as the classical lagrangian of the particle in the inverted potential Fig.~\ref{inverted_potential}.

One can solve for the minimum of the action~\eqref{euclidean_action} by putting its first variation to zero.
The action is minimized by the trajectories $x(\tau)\equiv$ corresponding to the particle sitting in either of the minima of the double-well potential. However, there are also additional imaginary-time-dependent solutions
\begin{equation}
x(\tau) =\pm \tanh \left( \tau -\tau _{0}\right),  \label{02}
\end{equation}%
where $\tau _{0}$ is an arbitrary time.
These solutions are usually called `instantons' or antiinstantons depending
on the sign. They describe the classical trajectories connecting the two ``humps'' of the inverted potential. In
the Hamiltonian language, this new solutions correspond to the tunnelling
between the two ``vacua'' leading to the splitting of the ground state.
Besides the single instanton or antiinstanton solutions, there are also the solutions consisting of multiple instantons and antiinstantons chained together.

The action at stationary solutions $x(\tau)\equiv \pm1$ equals zero. At the same time, its value at the instanton solutions is higher: $S_\mathrm{inst} = 4/3$. In the condensed matter setting, we should scale the action~\eqref{euclidean_action} by the volume of system $V$. As the result, the contribution of instanton solutions is exponentially small in the thermodynamic limit.
We should note that this picture stays valid if we consider any potential with the shape similar to the one displayed in Fig.~\eqref{euclidean_action}: there would be instanton solutions, however their action would be higher than that for the time-independent solutions.



It is interesting to note that the study of instanton physics has been pushed forward in the field of QCD (see e.g. \cite{shuryak})
in order to understand the structure of the ``ground'' state in that theory.
The situation there is quite to similar to what we have just discussed, namely, there are many different minima of the Euclidean action corresponding to the different vacua of the theory, and there are instanton configurations connecting these vacua.
However, the number of vacua is infinite, and the exponential supression of the instanton configurations caused by the increased action is compensated by the increased phase space factor.
As a result, the ``ground'' state of the system is described by a non-trivial combination of instantons, which is referred to in the field of QCD as ``instanton fluid''.

In view of these interesting developments,
it is natural to ask whether it is possible to formulate a model in the setting of Condensed Matter Physics,
which would admit a thermodynamically stable state described by an order parameter consisting of a system of instantons and antiinstantons?
Of course, one should expect that such a model should be consideraly more complicated than the simple toy model given by Eq.~\eqref{euclidean_action}. Still, the question remains the same.



In this paper, we suggest a new model that allows us to obtain in some region of parameters the ground
state with the imaginary-time-dependent order parameter. This order
parameter can be visualized as a lattice of alternating instantons and
antiinstantons, and therefore we coin for this phase the name ``instanton crystal''.
The model contains both interacting fermions and boson modes.
It does not contain any infinite or long-range interactions,
thus this model is in all respects rather conventional for condensed matter physics.
Using the mean-field approximation, we solve this model both analytically (in certain region of parameters) and numerically.
The study reveals a competition between the phase described by the static time-independent order parameter and the instanton crystal phase with a transition between the two phases.


In spite of the popularity of the instanton physics in QCD, only few works
have been done in the past on the investigation of a possibility of
non-perturbative effects in imaginary-time representation in condensed
matter physics. To the best of our knowledge, this problem has been first
attacked in Refs.°\cite{mukhin,mukhin1,mukhin2,mukhin2019} using a two-band
model of interacting fermions. A solution with the chain of instantons and
antiinstantons has been obtained. Unfortunately, it has finally been
concluded that in this model the free energy for the
imaginary-time-dependent configuration was always higher that for the static
configuration. A similar model with a coordinate instead of the imaginary
time $\tau \ $had been used long ago as 1D models of polymers, and a
solution with the chain of kinks and antikinks (spatial analogue of instantons and antiinstantons) had been obtained.
Ironically, it this case the energy of the ``instantons'' in the coordinate
space could be lower than that for the homogeneous solution.

Non-perturbative quantum dynamic effects have been studied in Ref. \cite%
{galitski} using the imaginary time representation.
Also, instanton-antiinstanton solutions appear in the studies of non-equilibrium systems \cite{spivak,barankov,altshuler,altshuler1,dzero}.

Recently, one more attempt has been undertaken \cite{efetov2019} to obtain
the instanton crystal. In that work, an additional interaction term has been
added to the previous model of Refs.~\cite{mukhin,mukhin1,mukhin2,mukhin2019}
with the hope that it might make the free energy of the instanton crystal
lower than the static state. Indeed, the presence of this new term reduced
the free energy but using the perturbation theory could not help proving the
existence of the instanton crystal. Many other guesses remained just guesses
due to the technical difficulties.

In the present paper, we modify the previous models by introducing bosonic current-like modes which are coupled to the fermions.
%
In comparison to the previous work~\cite{efetov2019}, we formulate an effective numerical scheme,
which enables us to solve the mean-field equations in the general case.
Moreover, in a certain region of
parameters the analytical treatment of the model is feasible. Results of both the
numerical and analytical study allow us to conclude that the instanton
crystal can exist in the thermodynamic equilibrium. Actually, this work is
the beginning of a systematic study of properties of the instanton crystal phase.

The paper is organized as follows.
In Section~\ref{model} we introduce the model without discussing its origin.
This is because we hope that the model is rather general. In Section~\ref{MF}
we minimize the effective Lagrangian and derive the mean-field equations.
In Section~\ref{analytics}, we solve the mean-field equations in a certain region of
the parameters of the model and calculate the free energy, demonstrating the possibility of the instanton crystal phase.
In Section~\ref{numerics} we describe the numerical scheme for the solution of mean-field equations,
which we use in Section~\ref{numerical_analysis} to conduct a detailed numerical investigation of the general case.
In Section~\ref{origin} we discuss a possible origin of the model.
Finally, in Section~\ref{outlook}, we discuss the results obtained and the perspectives of the future studies.
The appendices contain technical details of the calculations.

\section{General model for the imaginary-time crystal.\label{model}}

\subsection{Hamiltonian of the model.}

In this section, we formulate a rather general macroscopic model of
interacting fermions and bosons without going into details of its possible
origin. The latter will be done in Sec.~\ref{origin} but here we simply
introduce the general Hamiltonian $\hat{H}$ and discuss its structure.

The total Hamiltonian $\hat{H}$ of the model 
consists of three parts:%
\begin{equation}
\hat{H}=\hat{H}_{\mathrm{0}}+\hat{H}_{\mathrm{int}}+\hat{H}_{\mathrm{B}}.
\label{a2}
\end{equation}%
In Eq.~\eqref{a2}, Hamiltonian $\hat{H}_{\mathrm{0}}$ stands for a system of
non-interacting fermions
\begin{equation}
\hat{H}_{\mathrm{0}}=\sum_{\mathbf{p}}c_{\mathbf{p}}^{\dagger}\left(
\varepsilon _{\mathbf{p}}^{+}\check{\mathrm{I}}+\varepsilon _{\mathbf{p}}^{-}%
\check{\Sigma} _{3}\right) c_{\mathbf{p}}.  \label{a3}
\end{equation}%
These fermions live in two bands $1$ and $2$. Four-component vectors
\begin{equation}
c_{\mathbf{p}}=\left( c_{1\mathbf{p}}^{1},c_{1\mathbf{p}}^{2},c_{2\mathbf{p}%
}^{1},c_{2\mathbf{p}}^{2}\right)  \label{a3a}
\end{equation}%
contain as components creation and destruction operators $c_{\alpha p}^{s}$
for the fermions from the bands $s=1,2$ with spin projections labeled by $\alpha =1,2$ (actually,
the spin variable $\alpha$ is not very important here). The vectors $c_{%
\mathbf{p}}^{+}$ are hermitian conjugated to $c_{\mathbf{p}}$ and contain
creation operators as components. The energies $\varepsilon^{\pm}_\mathbf{p}$
are expressed in terms of the spectra $\varepsilon_{1,2}\left( \mathbf{p}%
\right) $ in the bands $1,2$ as
\begin{equation}
\varepsilon^{\pm}_\mathbf{p} =\frac{1}{2}\left( \varepsilon _{1}\left(
\mathbf{p}\right) \pm \varepsilon _{2}\left( \mathbf{p}\right) \right).
\label{a4}
\end{equation}%
The operators $\check{\Sigma}_{i}$, $i=1,2,3$ are Pauli matrices acting in
the subspace of the bands $1$ and $2$, while $\check{\mathrm{I}}$ is the
identity operator acting in the same subspace.

The second term $\hat{H}_{\mathrm{int}}$ in Eq.~\eqref{a2} stands for the
interaction between the fermions from different bands
\begin{multline}
\hat{H}_{\mathrm{int}}=-\cfrac{U_0}{4V}\sum_{\mathbf{p}_{1},\mathbf{p}_{2},%
\mathbf{q}}\left( c_{\mathbf{p}_{1}}^{\dagger }\check{\Sigma}_{2}c_{\mathbf{p%
}_{1}+\mathbf{q}}\right) \left( c_{\mathbf{p}_{2}}^{\dagger }\check{\Sigma}%
_{2}c_{\mathbf{p}_{2}-\mathbf{q}}\right) + \\
\*+\cfrac{\tilde{U}_0}{4V}\sum_{\mathbf{p}_{1},\mathbf{p}_{2},\mathbf{q}%
}\left( c_{\mathbf{p}_{1}}^{\dagger }\check{\Sigma}_{1}c_{\mathbf{p}_{1}+%
\mathbf{q}}\right) \left( c_{\mathbf{p}_{2}}^{\dagger }\check{\Sigma}_{1}c_{%
\mathbf{p}_{2}-\mathbf{q}}\right) .  \label{a4a}
\end{multline}%
where $V$ is the volume of the system. The Hamiltonian $\hat{H}_{\mathrm{int}%
}$ contains contact attraction (first term) and repulsion (second term).
Actually, the first term in Eq. (\ref{a4a}) describes attraction of
fermionic currents, while the second one stands for repulsion of charges.
More information about the possible origin of the model and interpretation
of the terms in the Hamiltonian is given in Sec. \ref{origin}. It is worth
emphasizing that the Hamiltonian $\hat{H}_{\mathrm{int}}$ does not contain
any long-range interactions.
We should also note that a model with the Hamiltonian $\hat{H}%
_{\mathrm{0}}+$ $\hat{H}_{\mathrm{int}}$ was considered previously in Ref.
\cite{efetov2019}) in a form adopted to a direct use of the mean-field
theory.

The third term $\hat{H}_{\mathrm{B}}$ describes a system of current-like
modes labeled by different momenta $\mathbf{q}$.
\begin{equation}
H_{\mathrm{B}}=\sum_{\mathbf{q}}\left[ \cfrac{U_2}{4}(\hat{\mathcal{P}}_{%
\mathbf{q}}-\hat{A}_{\mathbf{q}})(\hat{\mathcal{P}}_{-\mathbf{q}}-\hat{A}_{-%
\mathbf{q}})+\cfrac{\omega_\mathbf{q}^2\hat{\BQ}_\mathbf{q}\hat{\BQ}_{-%
\mathbf{q}}}{U_2}\right] ,  \label{a4b}
\end{equation}%
Where $\hat{\mathcal{Q}}_{\mathbf{q}}$ and $\hat{\mathcal{P}}_{\mathbf{q}}$
are conjugated coordinates and momenta of these modes satisfying the
following relations
\begin{equation}
\hat{\mathcal{Q}}_{\mathbf{q}}^{\dagger }=\hat{\mathcal{Q}}_{-\mathbf{q}%
},\quad \hat{\mathcal{P}}_{\mathbf{q}}^{\dagger }=\hat{\mathcal{P}}_{-%
\mathbf{q}}
\end{equation}%
The modes could be, in principle, just phonons but the latter generate
extremly low currents with respect to the fermionic ones (of the order $m/M,$ where $%
m$ is the electron mass, while $M$ is of the order of atomic masses) that
cannot efficiently interact with fermionic currents.

The current-like modes are coupled to the vector potential $\hat{A}_{\mathbf{q}}$ created by the fermions:
\begin{equation}
\hat{A}_{\mathbf{q}}=\cfrac{1}{\sqrt{V}}\sum_{\mathbf{p}}c_{\mathbf{p}%
}^{\dagger }\check{\Sigma}_{2}c_{\mathbf{p}+\mathbf{q}}.  \label{a8}
\end{equation}%
For a typical electron-phonon interaction, the electrons are coupled to
coordinates of phonons. In contrast, in our case, the fermions are coupled
to the momenta of the modes. In other words, we include into consideration
current-current interaction. This is an unusual feature, and it is crucial
for our new results. However, we argue in Sec. \ref{origin} that the
existence of these modes and their interaction with the fermions is not
unrealistic.

We assume that all the coupling constants are not negative, namely
\begin{equation}
U_{\mathrm{0}}\geq 0,\quad \tilde{U}_{\mathrm{0}}\geq 0,\quad U_{\mathrm{2}%
}\geq 0.
\end{equation}%
As usual, in the limit of large volume $V\rightarrow +\infty $, one can
replace the sum over the momenta by integrals using the standard replacement
\begin{equation}
\sum_{\mathbf{p}}\left( ...\right) \rightarrow V\int \left( ...\right) \frac{%
d\mathbf{p}}{\left( 2\pi \right) ^{d}},  \label{a9}
\end{equation}%
($d$ is dimension), which allows one to see that $\hat{H}$ is proportional
to the volume $V$, as it should be.


In this paper we restrict ourselves to studying thermodynamic
real-time-independent properties of the model. 
In order to study them,
the partition function $\mathcal{Z}$ in grand canonical ensemble is introduced in the standard way
\begin{equation}
\mathcal{Z}=\mathrm{Tr\exp }\left[ -\frac{\hat{H}-\mu \hat{N}}{T}\right] ,
\label{a1}
\end{equation}%
where $\mu$ is chemical potential, and it is convenient in the following to absorb it into the definition of $\varepsilon_{1,2}(\mathbf{p})$.

The Hamiltonian $\hat{H},$ Eqs. (\ref{a2}-\ref{a8}), describes a system of
interacting fermions and bosons, and it does not
contain any long-range interactions.
We would also like to emphasize that neither the Hamiltonian $\hat{H}$ nor the partition function $\mathcal{Z}$
contain any time-dependence, be it real time or imaginary time.
As a consequence, there can be no doubts that the Hamiltonian $\hat{H}$ describes a rather convential system in thermodynamic equilibrium.

In principle, one could proceed with the analysis of the model using the operator formalism.
However, it is more convenient for our study to use the corresponding Lagrangian formulation based on rewriting the partition function in the form of the functional integral over commuting and anticommuting fields.
This way, the imaginary time also enters the picture.
At the same time, we found instructive to provide the explicit Hamiltonian of the system in operator formalism:
it helps to stress the fact that we are considering a system in thermodynamic equilibrium without any pumping or relaxation.

\subsection{Field theory for the model under consideration.}

In the Lagrangian formulation, the partition function $\mathcal{Z}$ can be
written in the form of a functional integral as%
\begin{equation}
\mathcal{Z}=\int \exp \left( -S\left[ \chi ,\chi ^{+},a\right] \right) D\chi
D\chi^{+}Da.  \label{zdef}
\end{equation}%
where the action $S\left[ \chi ,\chi ^{+},a\right] $ contains anticommuting
fermionic fields $\chi^s_\alpha(\tau)$, $\chi^{s+}_\alpha(\tau)$ and
commuting bosonic fields $a_\mathbf{q}(\tau)$ and reads
\begin{equation}
S\left[\chi ,\chi ^{+},a\right] =S_{\mathrm{0}}+S_{\mathrm{int}}+S_{\mathrm{B%
}}+S_{\mathrm{FB}}.  \label{adef:full}
\end{equation}%
In Eq. (\ref{adef:full}), the terms $S_{\mathrm{0}}$ and $S_{\mathrm{int}}$
correspond to the terms $\hat{H}_{\mathrm{0}}$ and $\hat{H}_{\mathrm{int}}$
in the Hamiltonian approach, Eq.~\eqref{a2}, respectively. At the same time,
the terms $S_{\mathrm{B}}$ and $S_{\mathrm{FB}}$ correspond to the term $%
\hat{H}_{\mathrm{B}}$. 

As usual, the imaginary time $\tau $ is defined for $0\leq \tau \leq \beta
\equiv 1/T$ where $T$ is the temperature. The fermionic fields $\chi
_{\alpha \mathbf{p}}^{s}\left( \tau \right) $, $\chi _{\alpha \mathbf{p}%
}^{s+}\left( \tau \right) $ obey standard antiperiodic boundary conditions
\begin{equation}
\chi _{\alpha \mathbf{p}}^{s}\left( \tau +\beta \right) =-\chi _{\alpha
\mathbf{p}}^{s}\left( \tau \right) ,\;\chi _{\alpha \mathbf{p}}^{s+}\left(
\tau +\beta \right) =-\chi _{\alpha \mathbf{p}}^{s+}\left( \tau \right) ,
\label{a17}
\end{equation}%
and have the structure identical to the vectors $c_{\mathbf{p}}$ and $c_{%
\mathbf{p}}^{\dagger }$. In contrast, the bosonic fields obey the periodic
boundary conditions
\begin{equation}
a_{\mathbf{q}}\left( \tau +\beta \right) =a_{\mathbf{q}}\left( \tau \right) .
\label{a18}
\end{equation}

The first term in Eq.~\eqref{adef:full}, $S_{\mathrm{0}}$, is the action of
non-interacting fermions
\begin{multline}
S_{\mathrm{0}}\left[ \chi ,\chi ^{+}\right] = \\
\* = \sum_{\mathbf{p}}\int\limits_{0}^{\beta}\chi _{\mathbf{p}}^{+}\left(
\tau \right) \left[ \left(\partial_{\tau}+\varepsilon^{+}_\mathbf{p}\right)%
\check{\mathrm{I}}+\varepsilon^{-}_\mathbf{p} \check{\Sigma}_{3}\right] \chi
_{\mathbf{p}}\left( \tau \right) d\tau.  \label{adef:fermion}
\end{multline}%
%
%
%
%
%
%
%
%
%
%

The interaction term $S_{\mathrm{int}}\left[ \chi ,\chi ^{+}\right] $ has
the form%
\begin{multline}
S_{\mathrm{int}}\left[ \chi ,\chi ^{+}\right] =-\cfrac{U_0}{4V}\sum_{\mathbf{%
p}_{1,}\mathbf{p}_{2},\mathbf{q}}\int\limits_{0}^{\beta }d\tau \left( \chi _{%
\mathbf{p}_{1}}^{+}\left( \tau \right) \check{\Sigma}_{2}\chi _{\mathbf{p}%
_{1}+\mathbf{q}}\left( \tau \right) \right) \times \\
\*\times \left( \chi _{\mathbf{p}_{2}}^{+}\left( \tau \right) \check{\Sigma}%
_{2}\chi _{\mathbf{p}_{2}-\mathbf{q}}\left( \tau \right) \right) + \\
\*+\frac{\tilde{U}_{\mathrm{0}}}{4V}\sum_{\mathbf{p}_{1,}\mathbf{p}_{2},%
\mathbf{q}}\int\limits_{0}^{\beta }d\tau \left( \chi _{\mathbf{p}%
_{1}}^{+}\left( \tau \right) \check{\Sigma}_{1}\chi _{\mathbf{p}_{1}+\mathbf{%
q}}\left( \tau \right) \right) \times \\
\*\times \left( \chi _{\mathbf{p}_{2}}^{+}\left( \tau \right) \check{\Sigma}%
_{1}\chi _{\mathbf{p}_{2}-\mathbf{q}}\left( \tau \right) \right) .
\label{adef:int}
\end{multline}%

The term $S_{\mathrm{B}}\left[ a\right] $ in Eq.(\ref{adef:full}) stands for
the action of the phonon-like modes%
\begin{equation}
S_{\mathrm{B}}\left[ a\right] =\frac{1}{U_{\mathrm{2}}}\sum_{\mathbf{q}%
}\int\limits_{0}^{\beta }\left[ \left\vert \frac{da_{\mathbf{q}}\left( \tau
\right) }{d\tau }\right\vert ^{2}+\omega _{\mathbf{q}}^{2}\left\vert a_{%
\mathbf{q}}\left( \tau \right) \right\vert ^{2}\right] d\tau .
\label{adef:boson}
\end{equation}%
where $a_{\mathbf{q}}(\tau )$ are complex fields satisfying
\begin{equation}
(a_{\mathbf{q}}(\tau ))^{\ast }=a_{-\mathbf{q}}(\tau ).
\label{real_constraint}
\end{equation}%
The fields $a_{\mathbf{q}}\left( \tau \right) $
correspond to the coordinates in the language of oscillator modes, and $da_{%
\mathbf{q}}\left( \tau \right) /d\tau $ correspond to their velocities.

Finally, the coupling of the fermions to the current-like modes is described
by the term $S_{\mathrm{FB}}\left[ \chi ,\chi ^{+}\right] $ in Eq. (\ref%
{adef:full}), which takes the form
\begin{multline}
S_{\mathrm{FB}}\left[ \chi ,\chi ^{+},a\right] = \\
\*=-\frac{1}{\sqrt{V}}\sum_{\mathbf{p,q}}\int\limits_{0}^{\beta }\left( \chi
_{\mathbf{p}}^{+}\left( \tau \right) \check{\Sigma}_{2}\chi _{\mathbf{p+q}%
}\left( \tau \right) \right) \frac{da_{\mathbf{q}}\left( \tau \right) }{%
d\tau }d\tau .  \label{adef:fb}
\end{multline}

The terms $S_{\mathrm{B}}\left[ a\right] $ and $S_{\mathrm{FB}}\left[ \chi
,\chi ^{+}\right] $ constitute together the imaginary time Lagrangian
corresponding to the Hamiltonian term $\hat{H}_{\mathrm{B}}$. The Lagrangian
formulation of the functional integral can be obtained writing the
phase-space functional integral corresponding to $\hat{H}_{\mathrm{B}}$ and
then integrating out the momenta.

In the model under consideration, the expression $\chi _{\mathbf{p}%
}^{+}\left( \tau \right) \check\Sigma _{2}\chi _{\mathbf{p+q}}\left( \tau
\right) $ describes a current (see Section~\ref{origin}). Therefore, one can
interprete the term $S_{\mathrm{FB}}\left[ \chi ,\chi ^{+},a\right] $ as the
interaction of fermionic and bosonic currents.


In principle, one can integrate out in Eq. (\ref{zdef}) either fermionic or
bosonic fields just in the beginning of calculations. In order to compare
the model described by Eqs.~\eqref{zdef} and~\eqref{adef:full} with models
studied in the previous works \cite{mukhin,mukhin1,mukhin2,mukhin2019,efetov2019}%
, it is helpful first to integrate out the bosonic fields. This leads to the
following representation of the partition function $\mathcal{Z}$,
\begin{equation}
\mathcal{Z}=\int \exp \left[ -S_{\mathrm{fermion}}\left[ \chi ,\chi ^{+}%
\right] \right] D\chi D\chi ^{+},  \label{zdef:eff}
\end{equation}%
where the effective fermionic action $S_{\mathrm{fermion}}\left[ \chi ,\chi
^{+}\right] $ takes the form
\begin{align}
& S_{\mathrm{fermion}}\left[ \chi ,\chi ^{+}\right] =S_{\mathrm{0}}\left[
\chi ,\chi ^{+}\right] +  \label{adef:eff} \\
& +\sum_{\mathbf{q},\mathbf{p}_{1},\mathbf{p}_{2}}\left[ \frac{\tilde{U}_{%
\mathrm{0}}}{4V}\int\limits_{0}^{\beta }d\tau \left( \chi _{\mathbf{p}%
_{1}}^{+}\left( \tau \right) \check{\Sigma}_{1}\chi _{\mathbf{p}_{1}+\mathbf{%
q}}\left( \tau \right) \right) \times \right.  \notag \\
& \times \left( \chi _{\mathbf{p}_{2}}^{+}\left( \tau \right) \check{\Sigma}%
_{1}\chi _{\mathbf{p}_{2}-\mathbf{q}}\left( \tau \right) \right) -\frac{1}{4V%
}\iint\limits_{0}^{\beta }d\tau d\tau ^{\prime }K\left( \tau -\tau ^{\prime
}|\omega _{\mathbf{q}}\right) \times  \notag \\
& \left. \vphantom{\int\limits_0^\beta}\times \left( \chi _{\mathbf{p}%
_{1}}^{+}\left( \tau \right) \check{\Sigma}_{2}\chi _{\mathbf{p}_{1}+\mathbf{%
q}}\left( \tau \right) \right) \left( \chi _{\mathbf{p}_{2}}^{+}\left( \tau
^{\prime }\right) \check{\Sigma}_{2}\chi _{\mathbf{p}_{2}-\mathbf{q}}\left(
\tau ^{\prime }\right) \right) \right] .  \notag
\end{align}%
The first and the second terms are the same as those in Eqs.~%
\eqref{adef:fermion} and~\eqref{adef:int}, while the third one contains both
attraction and repulsion due to a special form of the interaction kernel
\begin{multline}
K\left( \tau -\tau ^{\prime }|\omega _{\mathbf{q}}\right) =\left( U_{\mathrm{%
0}}+U_{\mathrm{2}}\right) \delta \left( \tau -\tau ^{\prime }\right) - \\
\*-U_{\mathrm{2}}K_{0}\left( \tau -\tau ^{\prime }|\omega _{\mathbf{q}%
}\right) ,  \label{kernel:def}
\end{multline}%
where
\begin{equation}
K_{0}\left( \tau -\tau ^{\prime }|\omega _{\mathbf{q}}\right) =\frac{\omega
_{\mathbf{q}}\cosh {\left[ \omega _{\mathbf{q}}\left( \frac{\beta }{2}%
-\left\vert \tau -\tau ^{\prime }\right\vert \right) \right] }}{2\sinh {%
\frac{\beta \omega _{\mathbf{q}}}{2}}}  \label{green0}
\end{equation}%
is the solution of the differential equation
\begin{equation}
\left[ -\frac{1}{\omega _{\mathbf{q}}^{2}}\frac{d^{2}}{d\tau ^{2}}+1\right]
K_{0}\left( \tau -\tau ^{\prime }|\omega _{\mathbf{q}}\right) =\delta \left(
\tau -\tau ^{\prime }\right) .  \label{green1}
\end{equation}

Putting in Eq.~\eqref{adef:eff} $U_{\mathrm{2}}=0$, one arrives at the model
considered in Ref.~\cite{efetov2019}. Putting in addition $\tilde{U}_{%
\mathrm{0}}=0$ one comes to the model studied in Refs.~\cite%
{mukhin,mukhin1,mukhin2,mukhin2019}. Both these models contain only the
interactions local in imaginary time. On the contrary, in our case, the
kernel $K\left( \tau -\tau ^{\prime }|\omega _{\mathbf{q}}\right) $ contains
the additional repulsion term which is non-local in imaginary time. This
term is new and very important for the present study. At this point, we
would also like to emphasize that the fermion-fermion interactions remain
short-ranged in the real space.

\section{Mean-Field Theory.\label{MF}}

\subsection{Mean-field action}

Starting with the effective fermionic action $S_\mathrm{fermion}\left[%
\chi,\chi^{+}\right]$, we could, in principle, develop the perturbation
expansion in the coupling constants of the interaction terms. However, the
framework of the perturbation theory is not suitable for studying the type
of problems we consider in this paper, which is typical for strongly
correlated systems. The most common alternative is to perform the change of
variables from the fermionic degrees of freedom to the bosonic collective
degrees of freedom. This transformation is convenient for analytical studies
and is absolutely neccessary for the numerical ones.

The standard way to facilitate this change of variables is to decouple the
interaction terms with the help of the Hubbard-Stratonovich transformation.
In our case, this procedure gives us a model of fermions interacting with
auxiliary bosonic fields $\mathfrak{b}_{\mathbf{q}}\left( \tau \right) $ and
$\mathfrak{b}_{1\mathbf{q}}\left( \tau \right) $ (corresponding to $\check{%
\Sigma}_{2}$ and $\check{\Sigma}_{1}$ fermionic terms respectively). Then,
the resulting integral over the fermionic fields $\chi ,\chi ^{+}$ can be
calculated exactly to obtain the final representation of the partition
function $\mathcal{Z}$ in the form of a functional integral over the fields $%
\mathfrak{b}_{\mathbf{q}}\left( \tau \right) $ and $\mathfrak{b}_{1\mathbf{q}%
}(\tau )$
\begin{equation}
\mathcal{Z}=\int \exp \left[ -S_{\mathrm{final}}\left[ \mathfrak{b},%
\mathfrak{b}_{1}\right] \right] D\mathfrak{b}D\mathfrak{b}_{1},  \label{a32}
\end{equation}%
where
\begin{eqnarray}
&&S_{\mathrm{final}}\left[ \mathfrak{b},\mathfrak{b}_{1}\right] =  \notag \\
&=&-\int\limits_{0}^{\beta }d\tau \sum_{\mathbf{q}}\Big[2\sum_{\mathbf{p}}%
\mathrm{tr}\left[ \ln \check{h}_{\mathbf{p,q}}\right] _{\tau ,\tau }-\tilde{U%
}_{\mathrm{0}}^{-1}\left\vert \mathfrak{b}_{1\mathbf{q}}\left( \tau \right)
\right\vert ^{2}\Big]+  \notag \\
&&+\sum_{\mathbf{q}}\iint\limits_{0}^{\beta }d\tau d\tau ^{\prime
}K^{-1}\left( \tau -\tau ^{\prime }|\omega _{\mathbf{q}}\right) \mathfrak{b}%
_{\mathbf{q}}\left( \tau \right) \mathfrak{b}_{-\mathbf{q}}\left( \tau
^{\prime }\right) .  \notag \\
&&  \label{a33}
\end{eqnarray}

In Eq. (\ref{a33}),
\begin{equation}
\check{h}_{\mathbf{p,q}}\left( \tau \right) =\check{h}_{0\mathbf{p}}\left(
\tau \right) +\check{h}_{\mathbf{p,q}}^{\mathrm{int}}\left( \tau \right) ,
\label{a22}
\end{equation}%
where
\begin{equation}
\check{h}_{0\mathbf{p}}\left( \tau \right) =\left( \partial _{\tau
}+\varepsilon _{\mathbf{p}}^{+}\right) \check{\mathrm{I}}+\varepsilon _{%
\mathbf{p}}^{-}\check{\Sigma}_{3},  \label{a22a}
\end{equation}%
and
\begin{eqnarray}
&&\check{h}_{\mathbf{p,q}}^{\mathrm{int}}\left( \tau \right)  \label{a34} \\
&=&-\frac{1}{\sqrt{V}}\sum_{\mathbf{q}}\Big[\mathfrak{b}_{\mathbf{q}}\left(
\tau \right) \check{\Sigma}_{2}+i\mathfrak{b}_{1\mathbf{q}}\left( \tau
\right) \check{\Sigma}_{1}\Big]e^{-\mathbf{q}\frac{d}{d\mathbf{p}}}.  \notag
\\
&&  \notag
\end{eqnarray}%
The fields $\mathfrak{b}_{\mathbf{q}}(\tau )$ and $\mathfrak{b}_{1\mathbf{q}%
}(\tau )$ in Eqs. (\ref{a32}-\ref{a34}) satisfy the constraint~analogous to
the one in Eq. (\eqref{real_constraint}) and obey the periodic boundary
conditions
\begin{equation}
\mathfrak{b}_{\mathbf{q}}\left( \tau +\beta \right) =\mathfrak{b}_{\mathbf{q}%
}\left( \tau \right) ,\quad \mathfrak{b}_{1\mathbf{q}}\left( \tau +\beta
\right) =\mathfrak{b}_{1\mathbf{q}}\left( \tau \right) .  \label{a31a}
\end{equation}%
The function $K^{-1}\left( \tau -\tau ^{\prime }|\omega _{\mathbf{q}}\right)
$ is the inverse of the interaction kernel $K(\tau -\tau ^{\prime }|\omega _{%
\mathbf{q}})$ and equals (see Appendix~\ref{inversion})
\begin{multline}
K^{-1}(\tau -\tau ^{\prime }|\omega _{\mathbf{q}})=\frac{1}{U_{\mathrm{0}%
}+U_{\mathrm{2}}}\delta (\tau -\tau ^{\prime })+ \\
\*+\frac{U_{\mathrm{2}}}{U_{\mathrm{0}}(U_{\mathrm{0}}+U_{\mathrm{2}})}%
K_{0}(\tau -\tau ^{\prime }|\tilde{\omega}_{\mathbf{q}}),
\label{kernel:def_inv}
\end{multline}%
where
\begin{equation}
\tilde{\omega}_{\mathbf{q}}=\omega _{\mathbf{q}}\sqrt{\frac{U_{\mathrm{0}}}{%
U_{\mathrm{0}}+U_{\mathrm{2}}}}.\label{modified_frequency}
\end{equation}

Function $K^{-1}\left( \tau -\tau ^{\prime }|\omega_\mathbf{q}\right) $ is
positive, which guarantees convergence of the integral over $\mathfrak{b}_{%
\mathbf{q}}$ in Eq.~(\ref{a32}).

\subsection{Minimum of the action and mean-field equations.}

Although Eqs. (\ref{a32}-\ref{a33}) can serve as a direct calculation
procedure, explicit computation of the functional integral (\ref{a32}) is
still difficult even numerically. This is rather typical problem in study of
strongly correlated systems. A standard way to overcome this problem is to
start with developing a proper mean-field approximation. In many cases, the
mean-field theory allows one to understand properties of new models and
figure out what are the possible states, phase transitions between the
states, etc. After these first properties are understood, one proceeds with
studying fluctuations. Very often they are not so important, at least
qualitatively, but it may happen that they lead to significant changes of
the mean-field picture. However, starting with the mean-field approximation
is the first step that is worth doing.


The mean-field approximation corresponds to the calculation of the
functional integral, Eq. (\ref{a32}), using the saddle point method. Within
this technique one should find the minimum of the action $S_{\mathrm{final}}%
\left[\mathfrak{b},\mathfrak{b}_{1}\right]$ and approximate the free energy $%
F$ as%
\begin{equation}
F=-T\ln Z=TS_{\mathrm{final}}^{\left( 0\right) },  \label{b1}
\end{equation}%
where $S_{\mathrm{final}}^{\left( 0\right) }$ is the action $S_{\mathrm{final%
}}\left[\mathfrak{b},\mathfrak{b}_{1}\right] $ at the minimum.

It is rather natural to seek the minimum of $S_{\mathrm{final}}\left[ b,b_{1}%
\right] $ at coordinate-independent fields. This means that one should take
the fields $b_{\mathbf{q}}\left( \tau \right) $, $b_{1\mathbf{q}}\left( \tau
\right) $ at $\mathbf{q}=0$
\begin{equation}
\mathfrak{b}_{\mathbf{q}=0}\left( \tau \right) =\sqrt{V}b\left( \tau \right)
,\text{ }\mathfrak{b}_{1,\mathbf{q}=0}\left( \tau \right) =\sqrt{V}%
b_{1}\left( \tau \right) ,  \label{b2}
\end{equation}%
The proportionality of $\mathfrak{b}_{\mathbf{q}=0}\left( \tau \right) $ and
$\mathfrak{b}_{1,\mathbf{q}=0}\left( \tau \right) $ to $\sqrt{V}$ is typical
for condensate functions, and $b\left( \tau \right) $ and $b_{1}\left( \tau
\right) $ play the role of order parameters. In the Hamiltonian language one
can say that, below the phase transition temperature, a macroscopic number
of bosons is located at the state with $\mathbf{q=}0$. At the same time, the
fact that $b\left( \tau \right) $ and $b_{1}(\tau )$ may depend on $\tau $
signals about completely new phase transitions and thermodynamic states.

The fields at non-zero $\mathbf{q}$ correspond to the fluctuations around
the saddle point of the action. We will not consider them in this paper.
Neglecting the fluctuations, one can introduce the free energy functional
\begin{align}
&\mathcal{F}[b(\tau), b_1(\tau)] \equiv TS_\mathrm{final}[\sqrt{V}\delta_{%
\mathbf{q},0}b(\tau), \sqrt{V}\delta_{\mathbf{q},0}b_1(\tau)]=  \notag \\
&{= -TV\int\limits_0^\beta d\tau\left[ 2\int\frac{d\mathbf{p}}{(2\pi)^2}
\mathrm{tr}\left[\ln{\check h_\mathbf{p}}\right]_{\tau,\tau} - \tilde{U}%
_0^{-1} b_1^2(\tau) \right]} +  \notag \\
&{+ TV\iint\limits_0^\beta d\tau d\tau^\prime
K^{-1}(\tau-\tau^\prime|\omega_0) b(\tau) b(\tau^\prime).}
\label{fenergy_functional}
\end{align}
Here,
\begin{equation}
\check h_\mathbf{p}(\tau) = \check h_{0\mathbf{p}} + \check h_\mathbf{p}^%
\mathrm{int}(\tau),
\end{equation}
where $\check h_{0\mathbf{p}}$ is determined by Eq.~\eqref{a22a} and
\begin{equation}
\check{h}_{\mathbf{p}}^{\mathrm{int}}\left( \tau \right) =-b\left( \tau
\right) \check\Sigma _{2}-ib_{1}\left( \tau \right) \check\Sigma _{1}.
\label{b8}
\end{equation}

The equations for the minimum of the free energy functional can obtained by
putting to zero its first variation
\begin{multline}
\int_{0}^{\beta }d\tau ^{\prime }K^{-1}\left( \tau -\tau ^{\prime }|\omega
_{0}\right) b\left( \tau ^{\prime }\right) = \\
\*=\int \frac{d\mathbf{p}}{\left( 2\pi \right) ^{d}}\mathrm{tr}\left[ \check{%
\Sigma}_{2}\check{G}_{\mathbf{p}}\left( \tau ,\tau \right) \right] ,
\label{b3}
\end{multline}%
\begin{equation}
b_{1}\left( \tau \right) =i\tilde{U}_{\mathrm{0}}\int \frac{d\mathbf{p}}{%
\left( 2\pi \right) ^{d}}\mathrm{tr}\left[ \check{\Sigma}_{1}\check{G}_{%
\mathbf{p}}\left( \tau ,\tau \right) \right] ,  \label{b4}
\end{equation}%
In Eqs.~\eqref{b3} and~\eqref{b4}, the Green function $\check{G}_{\mathbf{p}%
}\left( \tau ,\tau ^{\prime }\right) $ satisfies the following equation%
\begin{equation}
\check{h}_{\mathbf{p}}\left( \tau \right) \check{G}_{\mathbf{p}}\left( \tau
,\tau ^{\prime }\right) =-\delta \left( \tau -\tau ^{\prime }\right) .
\label{b6}
\end{equation}%
We should note that putting $U_{\mathrm{2}}=0$ in Eqs.~\eqref{b3} and~%
\eqref{b4}, we come to the mean-field equations of Ref.~\cite{efetov2019}.

Equations~\eqref{b3} and~\eqref{b8} admit both the
imaginary-time-independent and the time-dependent solutions. In this paper,
we are going to show that in some region of parameters, an imaginary
time-dependent solution is energetically more favorable. However, in order
to determine the favorable configuration, one not only needs to obtain the
different solutions of the mean-field equations but to also calculate the
corresponding values of the free energy functional and compare them with
each other. As it appears, this is quite a non-trivial task. In Section~\ref%
{analytics}, we study analytically the limiting case of $U_{2}\ll U_{0}$.
The general case can only be tackled numerically. Thus, in Section~\ref%
{numerics}, we formulate a suitable computational scheme to treat the case
of general parameters, while the applications of the scheme are discussed in Section~\ref{numerical_analysis}.

It is worth mentioning that the origin of the interesting physics is the
existence of the non-zero imaginary-time-dependent order parameter $b(\tau )$%
. At the same time, the field $b_{1}(\tau )$ plays rather a supporting role
helping to decrease the free energy for the time-dependent configurations of
the field $b(\tau )$.
As a consequence, we are going to neglect the field $b_1(\tau)$ in our analysis of the model to simplify the calculations.
On the other hand, the inclusion of this field might be important for realistic description of experiments.



\section{Analytical study in the limit $U_{\mathrm{2}}\ll U_{\mathrm{0}}$, $%
\tilde{U}_{\mathrm{0}}=0$.\label{analytics}}

In the case $\tilde{U}_{\mathrm{0}}=0$, $U_{\mathrm{2}}=0$, the exact
solutions of Eq.~\eqref{b3} are known. However, the free energy of the
imaginary-time-independent configuration happens to be the lowest. In the
limit $U_{\mathrm{2}}\ll U_{\mathrm{0}}$, we can treat the non-local term in
the inverse kernel $K^{-1}(\tau -\tau ^{\prime }|\omega _{0})$ (see Eq.~%
\eqref{kernel:def_inv}) perturbatively making expansion in $U_{\mathrm{2}%
}/U_{\mathrm{0}}$. As a result of these procedure, one can obtain
analytically the first order corrections to the exact solutions for $U_{%
\mathrm{2}}=0$ as well as to the corresponding free energies. As we will
show, one can identify the region of parameters for which the free energy
of the time-independent configuration gets pushed above the corresponding
energy of the time-dependent configurations.

\subsection{Analysis of the case $U_{\mathrm{2}}=0$, $\tilde{U}_{\mathrm{0}%
}=0$.}

If one puts $\tilde{U}_{\mathrm{0}}=0$ and $U_{\mathrm{2}}=0$ in Eq.~%
\eqref{b3}, it gets transformed into
\begin{equation}
\cfrac{b(\tau)}{U_0}=-\int \cfrac{d\mathbf{p}}{(2\pi)^2}\mathrm{tr}\left[
\check{\Sigma}_{2}\check{G}_{\mathbf{p}}(\tau ,\tau )\right] ,
\label{lgap:local}
\end{equation}%
where
\begin{equation}
\left[ \check{h}_{0\mathbf{p}}(\tau )-b(\tau )\check{\Sigma}_{2}\right]
\check{G}_{\mathbf{p}}(\tau ,\tau ^{\prime })=-\check{\mathrm{I}}\delta
(\tau -\tau ^{\prime }).  \label{gfunction:simplified}
\end{equation}%
Equations~\eqref{lgap:local} and~\eqref{gfunction:simplified} has static
solutions $b(\tau )\equiv \pm \gamma _{T}$. The parameter $\gamma _{T}$ here
is determined by the self-consistency equation
\begin{multline}
\cfrac{2}{U_0}= \\
\*=\int \cfrac{d\mathbf{p}}{(2\pi)^2}\cfrac{\tanh{\frac{\beta(\kappa_%
\mathbf{p}^{(0)} + \varepsilon^+_\mathbf{p})}{2}} +
\tanh{\frac{\beta(\kappa_\mathbf{p}^{(0)} -
\varepsilon^+_\mathbf{p})}{2}}}{\kappa_\mathbf{p}^{(0)}}.
\label{self_consistency:static}
\end{multline}%
with
\begin{equation}
\kappa _{\mathbf{p}}^{(0)}=\sqrt{(\varepsilon _{\mathbf{p}}^{-})^{2}+\gamma
_{T}^{2}}
\end{equation}%
Of course, there is also a trivial solution $b(\tau )\equiv 0$, however we
assume that the parameters of the system are such that a non-trivial static
solution exists and is more energetically favorable than the trivial one.

The interesting fact is that, besides the static solutions $b(\tau )\equiv
\pm \gamma _{T}$, there is also a whole family of oscillating solutions
consisting of the instanton-antiinstanton pairs (see \cite%
{mukhin,mukhin1,mukhin2,mukhin2019,efetov2019}), bouncing back and forth
between the two static solutions $\pm \gamma _{T}$. These class of solutions
can be written exactly in terms of Jacobi elliptic function $\sn{(x|k)}$:
\begin{equation}
b(\tau )=k\gamma \sn{(\gamma(\tau-\tau_0)| k)}.  \label{b20}
\end{equation}%
For a solution corresponding to $m$ instanton-antiinstanton pairs, the
parameters $k$ and $\gamma $ should satisfy the system of equations
\begin{equation}
\beta =m\times \cfrac{4K(k)}{\gamma},  \label{period_equation}
\end{equation}%
\begin{multline}
\cfrac{2}{U_0}= \\
\*=\int \cfrac{d\mathbf{p}}{(2\pi)^2}\cfrac{|\varepsilon^-_\mathbf{p}|\left[%
\tanh{\frac{\beta(\kappa_\mathbf{p} + \varepsilon^+)}{2} +
\tanh{\frac{\beta(\kappa_\mathbf{p} - \varepsilon^+)}{2}}} \right]}
{\sqrt{\left((\varepsilon_\mathbf{p}^-)^2 +
\gamma^2\frac{(1-k)^2}{4}\right)\left((\varepsilon_\mathbf{p}^-)^2 +
\gamma^2\frac{(1+k)^2}{4}\right)}}.  \label{self_consistency}
\end{multline}%
The parameter $\kappa _{\mathbf{p}}$ is given by
\begin{equation}
\kappa _{\mathbf{p}}=|\varepsilon _{\mathbf{p}}^{-}|\sqrt{%
\cfrac{\left((\varepsilon_\mathbf{p}^-)^2 +
\gamma^2\frac{(1-k)^2}{4}\right)}{\left((\varepsilon_\mathbf{p}^-)^2 +
\gamma^2\frac{(1+k)^2}{4}\right)}}\cfrac{\Pi(n, \tilde{k})}{K(\tilde{k})},
\label{kappa}
\end{equation}%
where
\begin{equation}
n=\cfrac{\gamma^2k}{(\varepsilon_\mathbf{p}^-)^2 + \gamma^2\frac{(1+k)^2}{4}}%
,\quad \tilde{k}=\cfrac{2\sqrt{k}}{1+k}.  \label{n_ktilde}
\end{equation}%
In Eqs.~\eqref{period_equation} and~\eqref{self_consistency}, $K(k)$ is the
complete elliptic integral of the first kind, while $\Pi (n,k)$ is the
complete elliptic integral of the third kind (see, for example, \cite%
{whittaker,as} to read more about elliptic integrals and elliptic
functions). The equation~\eqref{period_equation} tells that the integer
number of instanton-antiinstanton pairs should fit onto the interval $%
[0,\beta ]$: one of the periods of $\sn(x|\tau )$ is $4K(k)$. Equation~%
\eqref{self_consistency} obtained from the condition of the minimum of the
free energy is actually the self-consistency equation in the mean-field
theory. In the case of large periods of instanton-antiinstanton pairs for
which $k\rightarrow 1$, Eq.~\eqref{self_consistency} simplifies into Eq.~%
\eqref{self_consistency:static}.

The solutions in the form of elliptic functions were used in \cite{mukhin,mukhin1, mukhin2, mukhin2019, efetov2019}.
For the convenience of the reader, we outline the derivation of the form of the solutions of Eq.~\eqref{lgap:local} and the
derivation of Eqs.~\eqref{self_consistency}, \eqref{kappa} and~%
\eqref{n_ktilde} in the Supplementary Material \cite{supplement}.



\subsection{Expansion in small $U_{\mathrm{2}}/U_{\mathrm{0}}$.}

Since we treat the non-local part of the kernel $K^{-1}(\tau -\tau ^{\prime
}|\omega _{0})$ as a perturbation, it is convenient to separate the
corresponding term in the free energy functional, Eq.~%
\eqref{fenergy_functional}. In the limit $U_{\mathrm{2}}/U_{\mathrm{0}}\ll 1$%
, we can neglect $U_{\mathrm{2}}$ when it appears in combination $U_{\mathrm{%
0}}+U_{\mathrm{2}}$ and write
\begin{multline}
\mathcal{F}[b(\tau )]=\mathcal{F}_{0}[b(\tau )]+ \\
\*+\cfrac{U_2}{U_0}\times TU_{\mathrm{0}}^{-1}\int_{0}^{\beta }d\tau
^{\prime }K_{0}(\tau -\tau ^{\prime }|\omega _{0})b(\tau )b(\tau ^{\prime }).
\label{fenergy_separated}
\end{multline}%
The perturbative expansion can be obtained if we substitute the ansatz
\begin{equation}
b(\tau )=b^{(0)}(\tau )+\frac{U_{\mathrm{2}}}{U_{\mathrm{0}}}b^{(1)}(\tau
)+\dotsb  \label{perturb_ansatz}
\end{equation}%
into the gap equation
\begin{equation}
\cfrac{\delta \BF[b(\tau)]}{\delta b(\tau)}=0.
\end{equation}%
In Eq.~\eqref{perturb_ansatz}, $b^{(0)}(\tau )$ is one of the solutions of
Eq.~\eqref{lgap:local} which is equivalent to $\delta \mathcal{F}_{0}[b(\tau
)]/\delta b(\tau )=0$.

Analogously, the corrections to the free energy can be obtained substituting
the ansatz~\eqref{perturb_ansatz} into Eq.~\eqref{fenergy_separated}. The
first order correction in the expansion of $b\left( \tau \right) $, Eq. (\ref%
{perturb_ansatz}), does not contribute in the first order to $\mathcal{F}%
_{0}[b(\tau )]$, because the $b^{\left( 0\right) }\left( \tau \right) $ is
obtained from the condition of the minimum of $\mathcal{F}_{0}[b(\tau )]$.
Then, the first order correction to $\mathcal{F}[b(\tau )]$ comes only from
the second term in Eq. (\ref{fenergy_separated}). All this means that, in
the first order in $U_{\mathrm{2}}/U_{\mathrm{0}}$, one can calculate $%
\mathcal{F}[b(\tau )]$ by simply inserting $b^{\left( 0\right) }\left( \tau
\right) $, Eq. (\ref{b20}), into both the terms in Eq. (\ref%
{fenergy_separated})$.$

So, we write the free energy $F$ in the form
\begin{equation}
F=\mathcal{F}[b^{(0)}(\tau )].  \label{fenergy_order1}
\end{equation}

\subsection{Comparison of the free energies of the instanton-antiinstanton
configurations with the free energy of the static configuration.}

Let us consider the static configuration $b_\mathrm{static}(\tau)\equiv
\gamma_T$ and a configuration consisting of $m$ instaton-antiinstanton pairs
$b_\mathrm{inst}^{m,k}(\tau) = k \gamma \sn(\gamma\tau|k)$ (parameter $%
\gamma $ is fixed by the choice of the parameters $m$ and $k$ according to
Eqs.~\eqref{period_equation}, \eqref{self_consistency}, \eqref{kappa} and~%
\eqref{n_ktilde}). We will denote the corresponding free energies as $F_%
\mathrm{static}$ and $F_\mathrm{inst}^{m,k}$. Using Eq.~%
\eqref{fenergy_order1}, we can write
\begin{multline}
F_\mathrm{inst}^{m,k} - F_\mathrm{static} = \left[\mathcal{F}_0[b_\mathrm{%
inst}^{m,k}(\tau)] - \mathcal{F}_0[\gamma_T]\right] + \\
\* + \cfrac{T U_2}{U_0^2} \iint\limits_0^\beta d\tau d\tau^\prime
K_0(\tau-\tau^\prime|\omega_0)b_\mathrm{inst}^{m,k}(\tau)b_\mathrm{inst}%
^{m,k}(\tau^\prime) - \\
\* - \cfrac{T U_2\gamma_T^2}{U_0^2} \iint\limits_0^\beta d\tau d\tau^\prime
K_0(\tau-\tau^\prime|\omega_0).
\end{multline}

To simplify the calculations, we consider the limit $k\rightarrow 1$. In
this limit, instanton-antiinstanton configurations spend almost all the time
in the vicinities of the static configurations $\pm \gamma_{0}$.
Correspondingly, if one neglects the non-local interaction term in Eq.~%
\eqref{b3}, the difference between the action of the instanton-antiinstanton
configuration and the action of the static configuration is proportional to
the number of the instanton-antiinstanton pairs. Thus, we can write the
difference of the free energies without the non-local interaction as
\begin{equation}
\left[ \mathcal{F}_{0}[b_{\mathrm{inst}}^{m,k}(\tau )]-\mathcal{F}%
_{0}[\gamma _{T}]\right] =T\Delta S_{0}m=\cfrac{\gamma_0 \Delta S_0}{4K(k)}.
\end{equation}%
Here, we used Eq.~\eqref{period_equation} using the fact that one can take
the parameter $\gamma $ to be equal to $\gamma _{T}$ for $k\rightarrow 1$.
Since the limit $k\rightarrow 1$ also corresponds to the limit of the zero temperature,
we take $\gamma_0 = \gamma_{T=0} = \gamma_0$.
The constant $\Delta S_{0}$ is the action difference for a single
instanton-antiinstanton pair:
\begin{equation}
\Delta S_{0}=2\int \frac{d\mathbf{p}}{(2\pi )^{2}}\left[ \ln {\frac{1+\frac{%
\gamma _{T}}{\sqrt{(\varepsilon _{\mathbf{p}}^{-})^{2}+\gamma _{T}^{2}}}}{1-%
\frac{\gamma _{T}}{\sqrt{(\varepsilon _{\mathbf{p}}^{-})^{2}+\gamma _{T}^{2}}%
}}}-\cfrac{2\gamma_0}{(\varepsilon_\mathbf{p}^-)^2 + \gamma_0^2}\right] >0.
\end{equation}
The correction to the free energy of the static configuration is evaluated
to be
\begin{equation}
\cfrac{T U_2\gamma_0^2}{U_0^2} \iint\limits_0^\beta d\tau d\tau^\prime
K_0(\tau-\tau^\prime|\omega_0) = \cfrac{U_2\gamma_0^2}{U_0^2}.
\end{equation}
The correction to the free energy of the instanton-antiinstanton
configuration can be calculated in the limit $\omega _{0}/\gamma_0\ll 1/K(k)$ using the Fourier expansion for the Jacobi elliptic function (see Appendix~\ref{correction_integral_calculation}).
This gives us
\begin{multline}
\cfrac{T U_2}{U_0^2}\iint\limits_{0}^{\beta }d\tau d\tau ^{\prime
}K_{0}(\tau -\tau ^{\prime }|\omega _{0})b_{\mathrm{inst}}^{m,k}(\tau )b_{%
\mathrm{inst}}^{m,k}(\tau ^{\prime })= \\
\*=\cfrac{U_2\gamma_0^2}{U_0^2}\times \frac{1}{3}\left( \frac{K(k)\omega
_{0}}{\gamma _{T}}\right) ^{2}.\label{correction_iai}
\end{multline}

Combining everything together, we obtain
\begin{multline}
F_{\mathrm{inst}}^{m,k}-F_{\mathrm{static}}= \\
\*=\cfrac{\gamma_0 \Delta S_0}{4K(k)}-\cfrac{U_2\gamma_0^2}{U_0^2}\times %
\left[ 1-\frac{1}{3}\left( \frac{K(k)\omega _{0}}{\gamma _{T}}\right) ^{2}%
\right] .
\end{multline}%
In the limit $k\rightarrow 1$, the elliptic integral $K\left( k\right) $
diverges, $K(k)\propto -\ln \left( 1-k\right) $. (This corresponds to a
large period of the instanton-antiinstanton lattice). So, it is possible to
choose $k$ sufficiently close to unity so that
\begin{equation}
\cfrac{\gamma_0 \Delta S_0}{4K(k)}-\cfrac{U_2\gamma_0^2}{U_0^2}<0.
\end{equation}%
Then, if we keep $\omega _{0}$ sufficiently small, the term quadratic in $%
\omega _{0}$ cannot change the sign of the free energy. As a result, the
instanton-antiinstanton configuration can really be energetically more
favourable.

This striking result that already very small coupling constants $U_{\mathrm{2%
}}\ll U_{\mathrm{0}}$ can make the state with instatons more favorable is
based on the strong sensitivity of the second term in Eq. (\ref%
{fenergy_separated}) to whether the solution $b^{\left( 0\right) }$ is
static or consists of the instanton-antiinstanton pairs in the imaginary
time $\tau $. In the former case this term can be large, while in the latter
case its value can considerably be reduced.

So, we have demonstrated here analytically that there is a region of the
parameters of the model, where the instanton crystal exists. Calculations in
a more broad region, as well as phase transitions between the states can be
studied only numerically and this will be done in the next section.



\section{Numerical minimization of the free energy functional.\label{numerics}}


At $U_2 = 0$ and $\tilde{U}_0=0$, the solutions of Eq.~\eqref{b3} with the different number $m$ of instanton-antiinstanton pairs are topologically distinct.
As $U_2$ is gradually turned on, the topological classes are kept intact.
As a result, the solution with $m$ pairs gets deformed yet the period of the configuration $W = \beta/m$ is preserved.

Alternatively, we can access different solutions of Eq.~\eqref{b3} if we minimize the free energy functional~\eqref{fenergy_functional} in the classes of configurations corresponding to different fixed periods $W = \beta/m$.
In order to turn this recepy into a numerical scheme, we just need to formulate a suitable discretization of the expression~\eqref{fenergy_functional} for the free energy functional, and find a way to enforce the restriction on the period of configurations.

We should also note, that we are going to neglect the field $b_1(\tau)$ for simplicity.
However, the resulting numerical scheme can be easily adapted to take this extra field into account.

\subsection{Transformation of the free energy functional}

The free energy functional, Eq.~\eqref{fenergy_functional}, consists of two parts: one is the part which is purely quadratic in the fields $b(\tau)$; another part is the fermionic part which originates from the integration of the fermionic degrees of freedom.
Discretization of the quadratic part of the free energy functional is a rather straightforward matter.
On the contrary, it is impossible to directly discretize the fermionic part of the free energy as it is written in Eq.~\eqref{fenergy_functional}.
However, it can be recast into the form suitable for the numerical treatment by replacing the functional trace with the expression involving a time-ordered exponential of the energy operator $\check h_\mathbf{p}(\tau)$.
This can be achieved with the help of the standard trick in the field of Determinant Monte Carlo (see, for example, Ref.~\cite{blankenbecler-81}).
We describe in details the transformation of the fermionic part of the free energy functional in Appendix~\ref{el_transformation}.

Using the same regularization as in Appendix~\ref{el_transformation}, we write the quantity of interest in our numerical studies as
\begin{multline}
 \frac{\BF[b(\tau)] - \BF_\mathrm{ferm}[0]}{V} = \\ \* = \frac{\BF_\mathrm{quad}^\mathrm{loc}[b(\tau)]}{V} +
 \frac{\BF_\mathrm{quad}^\mathrm{nloc}[b(\tau)]}{V} + \frac{\BF_\mathrm{ferm}[b(\tau)] - \BF_\mathrm{ferm}[0]}{V},\label{fenergy}
\end{multline}
where
\begin{equation}
 \frac{\BF_\mathrm{quad}^\mathrm{loc}[b(\tau)]}{V} = T\int\limits_0^\beta d\tau \frac{b^2(\tau)}{U_0 + U_2},\label{fenergy_quad_loc}
\end{equation}
\begin{multline}
 \frac{\BF_\mathrm{quad}^\mathrm{nloc}[b(\tau)]}{V} = \\ \* = \frac{TU_2}{U_0(U_0+U_2)}\iint\limits_0^\beta d\tau d\tau^\prime K_0(\tau-\tau^\prime|\tilde\omega_0) b(\tau)b(\tau^\prime)\label{fenergy_quad_nloc}
\end{multline}
and (see Eq.~\eqref{fenergy_basis})
\begin{multline}
 \frac{\BF_\mathrm{ferm}[b(\tau)] - \BF_\mathrm{ferm}[0]}{V} =
  - 
 2T\int \cfrac{d\mathbf{p}}{(2\pi)^2}\times \\ \* \times
 \ln{\cfrac{2\cosh{\beta\varepsilon_\mathbf{p}^+} + \tr\left[\BT e^{-\int_0^\beta d\tau (\varepsilon^-_\mathbf{p}\check\Sigma_3 - b(\tau)\check\Sigma_2)}\right]}{2\left(\cosh{\beta\varepsilon_\mathbf{p}^+} + \cosh{\beta\varepsilon_\mathbf{p}^-}\right)}}.\label{fenergy_f}
\end{multline}

Let us consider configuration $b(\tau)$ with period $W$ such that $m$ periods of the configuration fit into the interval $[0,\beta]$. We shall rewrite Eqs.~(\ref{fenergy}-\ref{fenergy_f}) in such a way that only the dependence on values of $b(\tau)$ for $\tau\in [0,W)$ explicitly enters the equations. This allows us to fix the period constraint for the optimization procedure.
\begin{equation}
 \frac{\BF_\mathrm{quad}^\mathrm{loc}[b(\tau)]}{V} = \frac{1}{W} \int\limits_0^W d\tau \frac{b^2(\tau)}{U_0+U_2},\label{period_quad_loc}
\end{equation}
\begin{multline}
 \frac{\BF_\mathrm{quad}^\mathrm{nloc}[b(\tau)]}{V}
 = \\ \* =
 \frac{1}{W}\sum\limits_{k=0}^{m-1} \iint\limits_0^W d\tau d\tau^\prime K_0(\tau - \tau^\prime - kW|\tilde\omega_0) b(\tau)b(\tau^\prime)
 = \\ \* =
 \frac{1}{W}\iint\limits_0^W d\tau d\tau^\prime \tilde K_0(\tau-\tau^\prime|\tilde\omega_0) b(\tau) b(\tau^\prime),\label{period_quad_nloc}
\end{multline}
where
\begin{multline}
 \tilde K_0(\tau-\tau^\prime|\tilde\omega_0) = \frac{\tilde\omega_0\cosh{\left[\tilde\omega_0\left(\frac{W}{2} - |\tau - \tau^\prime|\right)\right]}}{2\sinh{\frac{W\tilde\omega_0}{2}}}.
\end{multline}
Notice that the expression for the averaged kernel $\tilde K_0(\tau-\tau^\prime|\tilde\omega_0)$ is identical to Eq.~\eqref{kernel:def_inv} for the kernel $K_0(\tau-\tau^\prime|\tilde\omega_0)$ with the only difference being that $\beta$ is replaced by $W$.

Finally, let us define
\begin{equation}
 \check U_\mathbf{p}(\tau_2, \tau_1) = \BT e^{-\int_{\tau_1}^{\tau_2}d\tau \left(\varepsilon_\mathbf{p}^-\check\Sigma_3 - b(\tau)\check\Sigma_2\right)}.
\end{equation}
With the help of this definition, we can rewrite Eq.~\eqref{fenergy_f} as
\begin{multline}
 \frac{\BF_\mathrm{ferm}[b(\tau)] - \BF_\mathrm{ferm}[0]}{V} = -\frac{2}{mW}\int \frac{d\mathbf{p}}{(2\pi)^2}\times \\ \* \times
\ln{\cfrac{2\cosh{\beta\varepsilon_\mathbf{p}^+} + \tr\left[\left(\check U_\mathbf{p}(W,0)\right)^m\right]}{2\left(\cosh{\beta\varepsilon_\mathbf{p}^+} + \cosh{\beta\varepsilon_\mathbf{p}^-}\right)}}.\label{period_f}
\end{multline}

\subsection{Discretization scheme.\label{scheme}}

The goal of discretizing the expression for the free energy is achieved if we replace the function $b(\tau)$ with its values at the discrete set of imaginary time points $b_i = b(\tau_i)$. For the discretization scheme with $N$ points, we are going to take $\tau_i = (i-1)\Delta\tau = (i-1)W/N$, where $i$ runs through integer values from $1$ up to $N$.
It is a straightforward task to write up the discretized versions of Eqs.~(\ref{period_quad_loc}-\ref{period_f}):
\begin{equation}
 \frac{\BF_\mathrm{quad}^\mathrm{loc}[b(\tau)]}{V}\rightarrow \frac{\BF_\mathrm{quad}^\mathrm{loc}[b_i]}{V} = \frac{\Delta\tau}{W}\sum\limits_{i=1}^N \frac{b_i^2}{U_0+U_2}.\label{discrete_quad_loc}
\end{equation}
\begin{equation}
 \frac{\BF_\mathrm{quad}^\mathrm{nloc}[b(\tau)]}{V} \rightarrow \frac{\BF_\mathrm{quad}^\mathrm{nloc}[b_i]}{V} = 
 \frac{\Delta\tau^2}{W}\sum\limits_{i,j=1}^N \tilde K_0(\tau_i-\tau_j|\tilde\omega_0) b_i b_j.\label{discrete_quad_nloc}
\end{equation}
\begin{multline}
 \frac{\BF_\mathrm{ferm}[b(\tau)] - \BF_\mathrm{ferm}[0]}{V}\rightarrow \frac{\BF_\mathrm{ferm}[b_i] - \BF_\mathrm{ferm}[0]}{V}
 = \\ \* = -\frac{2}{mW}\int \frac{d\mathbf{p}}{(2\pi)^2}
\ln{\cfrac{2\cosh{\beta\varepsilon_\mathbf{p}^+} + \tr\left[\left(\check U_\mathbf{p}[b_i]\right)^m\right]}{2\left(\cosh{\beta\varepsilon_\mathbf{p}^+} + \cosh{\beta\varepsilon_\mathbf{p}^-}\right)}},\label{discrete_f}
\end{multline}
where $\check U_\mathbf{p}[b_i]$ is the discrete approximation of the time-ordered exponential:
\begin{equation}
 \check{U}_\mathbf{p}(W,0)\rightarrow \check U_\mathbf{p}[b_i] = \prod\limits_{i = N}^1 e^{-\Delta\tau\left(\varepsilon_\mathbf{p}^-\check\Sigma_3 - b_i \check\Sigma_2\right)}.\label{discrete_texp}
\end{equation}
Substituting this expressions into Eq.~\eqref{fenergy}, we obtain the discretized version of the full free energy functional:
\begin{multline}
 \frac{\BF[b_i] - \BF_\mathrm{ferm}[0]}{V} = \\ \* = \frac{\BF_\mathrm{quad}^\mathrm{loc}[b_i]}{V} +
 \frac{\BF_\mathrm{quad}^\mathrm{nloc}[b_i]}{V} + \frac{\BF_\mathrm{ferm}[b_i] - \BF_\mathrm{ferm}[0]}{V}.
\end{multline}

To run the optimization procedure, we also need the formulas for the gradient of the free energy
\begin{equation}
 \frac{\partial}{\partial b_i}\left(\frac{\BF_\mathrm{quad}^\mathrm{loc}[b_i]}{V}\right) = \frac{2\Delta\tau}{W}\frac{b_i}{U_0+U_2}.\label{grad_quad_loc}
\end{equation}
\begin{equation}
 \frac{\partial}{\partial b_i}\left(\frac{\BF_\mathrm{quad}^\mathrm{nloc}[b_i]}{V}\right) = \frac{2\Delta\tau^2}{W}\sum\limits_{j = 1}^N \tilde K_0(\tau_i-\tau_j)b_j.\label{grad_quad_nloc}
\end{equation}
\begin{multline}
 \frac{\partial}{\partial b_i}\left(\frac{\BF_\mathrm{ferm}[b_i] - \BF_\mathrm{ferm}[0]}{V}\right) = \\ \* =
 -\frac{2}{W}\int \frac{d\mathbf{p}}{(2\pi)^2} \frac{\tr{\left[\left(\check U_\mathbf{p}[b_i]\right)^{m-1} \partial_{b_i}\check U_\mathbf{p}[b_i]\right]}}{2\cosh{\beta\varepsilon_\mathbf{p}^+} + \tr\left[\left(\check U_\mathbf{p}[b_i]\right)^m\right]},\label{grad_f}
\end{multline}
where
\begin{multline}
 \partial_{b_i}\check U_\mathbf{p}[b_i] = \prod\limits_{j = N}^{i+1} e^{-\Delta\tau\left(\varepsilon_\mathbf{p}^-\check\Sigma_3 - b_j \check\Sigma_2\right)} \times \\ \* \times \frac{\partial e^{-\Delta\tau\left(\varepsilon_\mathbf{p}^-\check\Sigma_3 - b_i \check\Sigma_2\right)}}{\partial b_i}\times \prod\limits_{j = i-1}^1 e^{-\Delta\tau\left(\varepsilon_\mathbf{p}^-\check\Sigma_3 - b_j \check\Sigma_2\right)}.\label{discrete_texp_prime}
\end{multline}

The explicit expressions for the matrices appearing in Eqs.~\eqref{discrete_texp} and~\eqref{discrete_texp_prime} are
\begin{multline}
 e^{-\Delta\tau\left(\varepsilon_\mathbf{p}^-\check\Sigma_3 - b_i\check\Sigma_2\right)} = \\ \* = \check{\mathrm{I}} \cosh{\kappa_{i\mathbf{p}}\Delta\tau} - \left(\varepsilon_\mathbf{p}^-\check\Sigma_3 - b_i\check\Sigma_2\right)\frac{\sinh{\kappa_{i\mathbf{p}}\Delta\tau}}{\kappa_{i\mathbf{p}}},
\end{multline}
\begin{multline}
 \frac{\partial e^{-\Delta\tau\left(\varepsilon_\mathbf{p}^-\check\Sigma_3 - b_i\check\Sigma_2\right)}}{\partial b_i} = 
 \left(\check\Sigma_2 + b_i\Delta\tau \check{\mathrm{I}}\right) \frac{\sinh{\kappa_{i\mathbf{p}}\Delta\tau}}{\kappa_{i\mathbf{p}}}
 - \\ \* -
 b_i\left(\varepsilon_\mathbf{p}^-\check\Sigma_3 - b_i\check\Sigma_2\right) \frac{\kappa_{i\mathbf{p}}\Delta\tau\cosh{\kappa_{i\mathbf{p}}\Delta\tau} - \sinh{\kappa_{i\mathbf{p}}\Delta\tau}}{\kappa_{i\mathbf{p}}^3}.
\end{multline}
Here, the parameter $\kappa_{i\mathbf{p}}$ is
\begin{equation}
 \kappa_{i\mathbf{p}} = \sqrt{(\varepsilon_\mathbf{p}^-)^2 + b_i^2}.
\end{equation}

The scheme we just introduced can be implemented in the programming language of the choice. Since the scheme provides the expressions both for the discretized free energy functional and its gradient, it can be plugged into any first order optimization algorithm.

The analytical solutions of Eq.~\eqref{b3} without the non-local part of $K_0(\tau-\tau^\prime|\tilde\omega_0)$ can be used as the initial conditions for the optimization procedure.
For fixed $W = \beta/m$, this requires to solve the system of equations~(\ref{period_equation}-\ref{n_ktilde}) to determine the parameters $k$ and $\gamma$. Since we neglect the non-local part and do not put $U_2 = 0$, one should replace $U_0$ by $U_0+U_2$ in Eq.~\eqref{self_consistency}. Then, the initial condition is defined as $b^{(0)}_i = k\gamma\sn(\gamma\tau_i|k)$.
Alternatively, one can use instead $b^{(0)}_i = k\gamma_T\sn(\gamma_T\tau_i|k)$ where $\gamma_T$ is the solution of the static gap equation~\eqref{self_consistency:static} (but with $U_0+U_2$ instead of $U_0$) and $k$ is determined from the condition $W = 4K(k)/\gamma_T$. In the end, both choices of the initial conditions lead to the same results of the optimization procedure.

\subsection{Variation of the scheme in the limit $T\rightarrow +0$.\label{zerotemp_scheme}}

The numerical scheme we introduced in the previous subsection can be adapted to treat the limiting case of zero temperature, which is equivalent to the limit $m\rightarrow +\infty$.
The only expressions that need to be adjusted are Eqs.~\eqref{discrete_f} and~\eqref{grad_f} which define the discretized version of the fermionic part of the free energy functional and its gradient.

Suppose that one calculates the matrix $\check{U}_\mathbf{p}[b_i]$ for some specific value of $\mathbf{p}$.
The diagonal decomposition of this matrix is given by
\begin{equation}
 \check U_\mathbf{p}[b_i] = \check{S}_\mathbf{p}
 \left(
    \begin{array}{cc}
     \lambda_{1\mathbf{p}} & 0\\
     0 & \lambda_{2\mathbf{p}}
     \end{array}
 \right)\check{S}^{-1}_\mathbf{p},
\end{equation}
where we assume that $\lambda_{1\mathbf{p}}$ is the eigenvalue with the largest absolute value.
Then, one can write Eq.~\eqref{discrete_f} in the limit $m\rightarrow +\infty$ as
\begin{multline}
  \frac{\BF_\mathrm{ferm}[b_i] - \BF_\mathrm{ferm}[0]}{V} = \\ \* = -2\int \frac{d\mathbf{p}}{(2\pi)^2}\
\left[\max{\left(\frac{\ln{\lambda_{1\mathbf{p}}}}{W}, |\varepsilon_\mathbf{p}^+|\right)} - \max{\left(|\varepsilon_\mathbf{p}^-|, |\varepsilon_\mathbf{p}^+|\right)}\right].
\end{multline}
Analogously, Eq.~\eqref{grad_f} transforms into
\begin{multline}
 \frac{\partial}{\partial b_i}\left(\frac{\BF_\mathrm{ferm}[b_i] - \BF_\mathrm{ferm}[0]}{V}\right) = 
 -\frac{2}{W}\int \frac{d\mathbf{p}}{(2\pi)^2} \times \\ \* \times\theta\left(\frac{\ln{\lambda_{1\mathbf{p}}}}{W} - |\varepsilon_\mathbf{p}^+|\right)\lambda_{1\mathbf{p}}^{-1}\left(\check S_\mathbf{p}^{-1}\partial_{b_i}\check{U}_\mathbf{p}[b_i]\check S_\mathbf{p}\right)_{1,1}.
\end{multline}
Here, $\theta(x)$ is the Heaviside function, while $\left(\check{A}\right)_{i,j}$ denotes the matrix element $i,j$ of the $2\times2$ matrix $\check{A}$.

\section{Numerical analysis of the model.\label{numerical_analysis}}

\subsection{Zero temperature.}

\begin{figure*}[t]
 \includegraphics[width=6.4in]{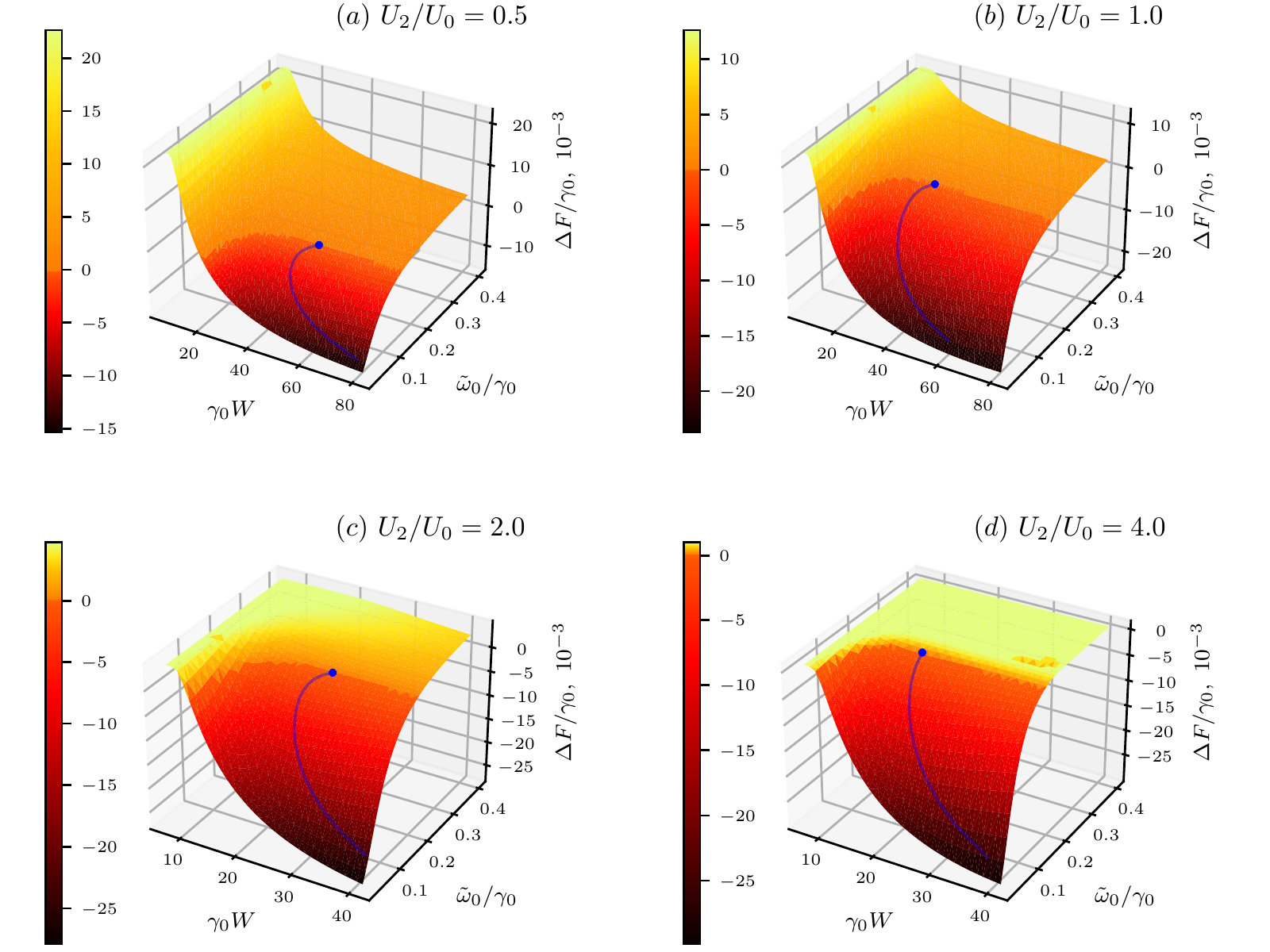}
 \caption{Dimensionless difference of the free energies $\Delta F/\gamma_0 = (F_\mathrm{instanton} - F_\mathrm{static})/\gamma_0$ at zero temperature as the function of the dimensionless period of the instanton lattice $\gamma_0 W$ and of the dimensionless parameter $\tilde \omega_0/\gamma_0$, characterizing the current-like mode. The four subplots correspond to the four different values of the ratio $U_2/U_0$: (a) $U_0/U_2 = 0.5$, (b)  $U_0/U_2 = 1.0$, (c) $U_2/U_0 = 2.0$ and (d) $U_2/U_0 = 4.0$.
 The parameters of the fermionic dispersion were fixed and their specific values are described in the main text.
 The blue curves show the minima of $\Delta F/\gamma_0$ at fixed value of $\tilde\omega_0/\gamma_0$ as the function of $\tilde\omega_0/\gamma_0$.
 The blue points at the end of the blue curves mark the transition between the instanton crystal phase and the static phase.}
 \label{energy_surfaces}
\end{figure*}

\begin{figure*}[t]
 \includegraphics[width=5.6in]{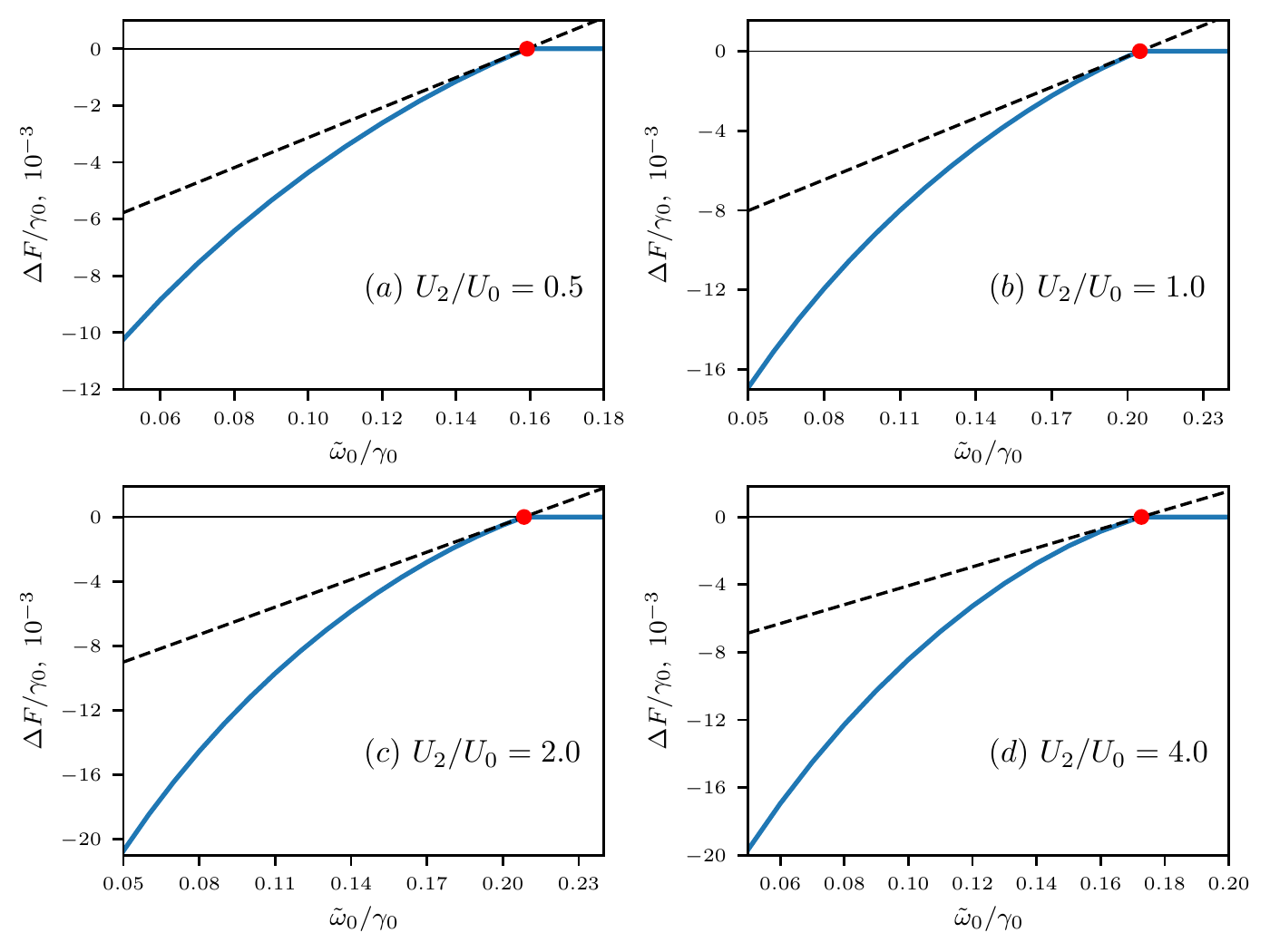}
 \caption{The slope $d(\Delta F/\gamma_0)/d(\tilde\omega_0/\gamma_0)$ (black dashed lines) of the dimensionless free energy difference $\Delta F/\gamma_0$ as the function of $\tilde\omega_0/\gamma_0$ (blue curves) in the instanton crystal phase at the transition point (red dots) to the static phase.
 The four subplots correspond to the same four different values of the ratio $U_2/U_0$ as in Fig.~\ref{energy_surfaces}: (a) $U_0/U_2 = 0.5$, (b)  $U_0/U_2 = 1.0$, (c) $U_2/U_0 = 2.0$ and (d) $U_2/U_0 = 4.0$.
 }
 \label{energy_derivative_jumps}
\end{figure*}

In order to perform the actual numerical simulations, one needs to specify the fermionic dispersion $\varepsilon_{1,2}(\mathbf{p})$ (see Eqs.~\eqref{a3} and~\eqref{a4}).
As we explain in Section~\ref{origin}, the model introduced in the present paper originates from the Spin-Fermion model with Overlapping Hotspots (SFMOHS) studied in references~\cite{volkov1,volkov2,volkov3}.
As a consequence, we have chosen the fermionic dispersion in the same form as it appears in SFMOHS:
\begin{equation}
 \varepsilon_1(\mathbf{p}) = \alpha p_x^2 - \beta p_y^2 -\mu,\quad \varepsilon_2(\mathbf{p}) = \alpha p_y^2 - \beta p_x^2 -\mu,
\end{equation}
where $\mu$ is the chemical potential.
We also introduce an energy cutoff $\Lambda$ limiting the width of dispersion:
\begin{equation}
 \frac{\alpha + \beta}{2}(p_x^2 + p_y^2) < \Lambda.
\end{equation}

We should note, that in all the computations we neglected the field $b_1(\tau)$, which is equivalent to setting $\tilde U_0 = 0$.

In Fig.~\ref{energy_surfaces}, we display the dimensionless difference between the free energies of the instanton-antiinstanton and of the static configurations $\Delta F/\gamma_0 = (F_\mathrm{instanton} - F_\mathrm{static})$ at zero temperature as the function of the dimensionless period of the instanton lattice $\gamma_0 W$ and of the dimensionless parameter $\tilde \omega_0/\gamma_0$ which corresponds to the modified frequency of the current-like mode (see Eq.~\eqref{modified_frequency}).
The energy scale $\gamma_0$, which we use to make the physical quantities dimensionless, is the solution of the static gap equation~\eqref{self_consistency:static} at zero temperature and in the abscence of the nonlocal repulsion term.
In addition to that, the same equation~\eqref{self_consistency:static} at zero temperature was used to determine the value of the dimensionless parameter $(U_0+U2)/\gamma_0$.
The four subplots of Fig.~\ref{energy_surfaces} correspond to the four different values of the ratio $U_2/U_0$: (a) $U_0/U_2 = 0.5$, (b)  $U_0/U_2 = 1.0$, (c) $U_2/U_0 = 2.0$ and (d) $U_2/U_0 = 4.0$. The parameters of the fermionic dispersion were kept fixed and their specific values were: $\alpha = \beta = 1.0$, $\Lambda/\gamma_0 = 1.0$, $\mu/\gamma_0 = 0.0$. The results were obtained using the zero-temperature variant of the numerical scheme described in subsection~\ref{zerotemp_scheme}.

In each of the subplots of Fig.~\eqref{energy_surfaces}, one can clearly identify the regions where the free energy of the instanton configurations becomes less than the free energy of the static configuration.
As a consequence, in these regions the instanton crystal phase should be the one which is thermodynamically stable.

In the instanton crystal phase, the actual period of the lattice is determined by the minimum of the free energy at fixed value of $\tilde \omega_0/\gamma_0$. In Fig.~\eqref{energy_surfaces}, the blue curves show the positions of the minima of $\Delta F/\gamma_0$ as the functions of $\tilde \omega_0/\gamma_0$. These minima were extracted by interpolation from the same data used to plot the surfaces. 
As the value of the parameter $\tilde \omega_0/\gamma_0$ grows, we observe the transition from the instanton crystal phase to the phase with imaginary-time-independent order parameter. In Fig.~\eqref{energy_surfaces}, this transition is marked by the blue points at the end of the blue curves.

Just below the transition, the period and the amplitude of the instanton lattice has finite values, as a result, the transition should be accompanied by an abrupt change in the order parameter. Thus, the transition must be of the first order.
In order to prove this point, we display in Fig.~\eqref{energy_derivative_jumps} the slope $d(\Delta F/\gamma_0)/d(\tilde\omega_0/\gamma_0) = d\Delta F/d\tilde\omega_0$ of the dimensionless free energy $\Delta F/\gamma_0$ as the function of $\tilde\omega_0/\gamma_0$ in the instanton crystal phase at the transition point for the same values of the ratio $U_2/U_0$ as in Fig.~\eqref{energy_derivative_jumps}. In each of the four cases, the slope has the finite value in the instanton crystal phase, while it is zero in the static phase, from which one can conclude that in each of the four cases the transition is accompanied by the jump in the first order derivative of the free energy $dF/d\tilde\omega_0 = d\Delta F/d\tilde\omega_0$.

\subsection{Finite temperatures}

\begin{figure*}[t]
 \includegraphics[width=5in]{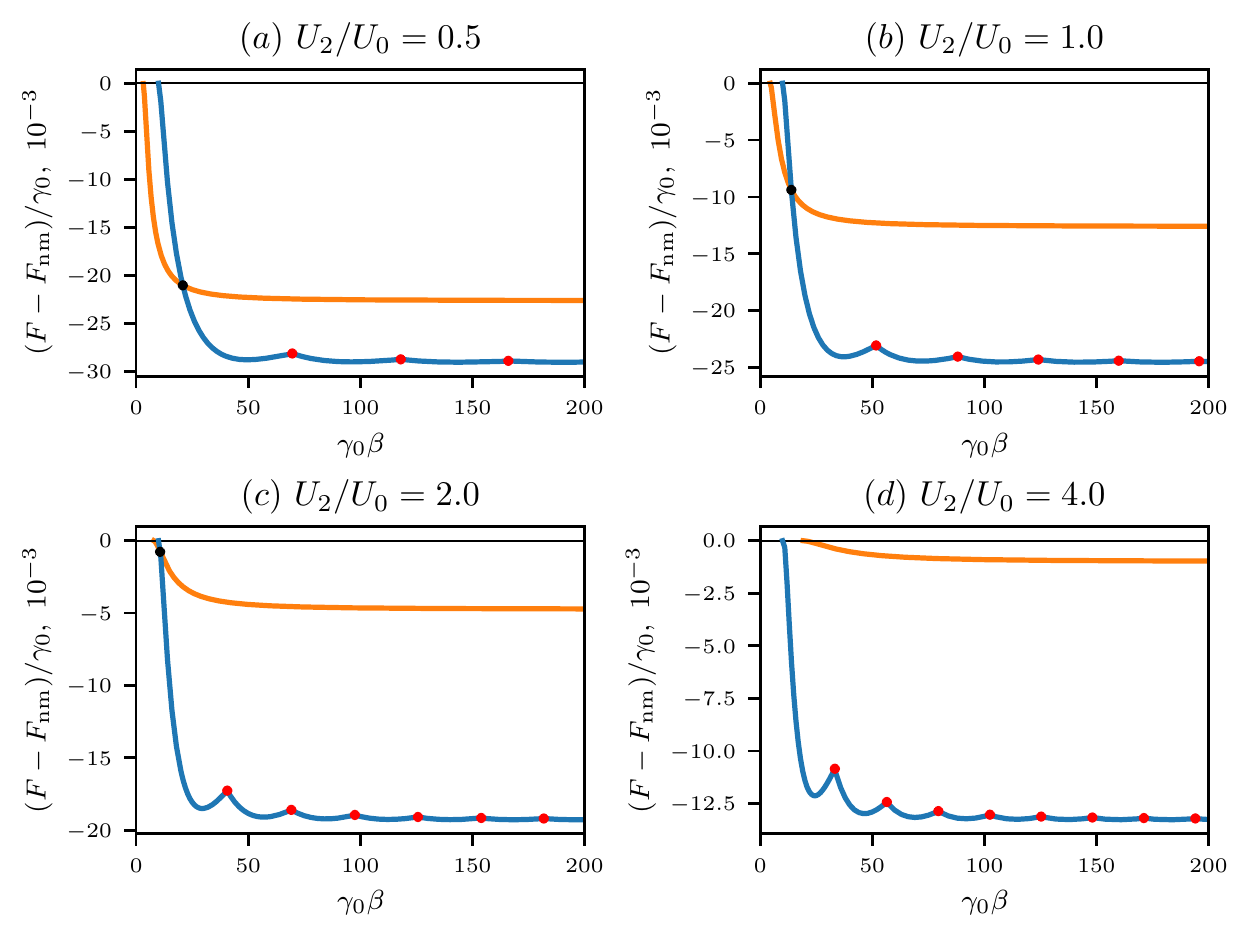}
 \caption{Comparison between the dimensionless free energy of the instanton crystal configuration (blue curves) and the dimensionless free energy of the static configuration (orange curves) as the functions of the dimensionless inverse temperature $\gamma_0 \beta$. The free energy of both of the types of configurations was determined with respect to the free energy of the normal metal configuration. As the inverse temperature grows, the instanton crystal undergoes first order transitions characterized by the change in the number $m$ of periods of the lattice by $1$. The points of these transitions are marked by the red dots.
 The four subplots correspond to the same four different values of the ratio $U_2/U_0$ as in Fig.~\ref{energy_surfaces}: (a) $U_0/U_2 = 0.5$, (b)  $U_0/U_2 = 1.0$, (c) $U_2/U_0 = 2.0$ and (d) $U_2/U_0 = 4.0$.
 }
 \label{temperature}
\end{figure*}

Besides the quantum phase transitions at zero temperature, it is also interesting to study how the model may enter the instanton crystal phase as the temperature is varied. Thus, we also calculated the dependence of the free energy of the instanton crystal configuration as the function of the inverse temperature.

The results of the calculations are presented in Fig.~\ref{temperature}.
There, we plot the dimensionless free energy of the instanton crystal configuration and the dimensionless free energy of the imaginary-time-independent configuration as the functions of the dimensionless inverse temperature $\gamma_0 \beta$. The free energies are determined with respect to the free energy of the normal metal configuration ($b(\tau)\equiv 0$).
For the calculations, we used the same parameters of the fermionic dispersion and the same set of ratios $U_2/U_0$ as we did for the calculations at zero temperature. In all the cases, the value of the parameter $\tilde\omega_0/\gamma_0$ was fixed: $\tilde\omega_0/\gamma_0 = 0.08$.  We should also emphasize that $\gamma_0$ and $(U_0+U_2)/\gamma_0$ were determined by the static gap equation~\eqref{self_consistency:static} at zero temperature, so that their values stayed constant as the temperature was varied.

For the cases $U_2/U_0 = 0.5$, $U_2/U_0 = 1.0$ and $U_2/U_0 = 2.0$ which correspond to the plots Fig.~\ref{temperature} (a,b,c), we observe that as the temperature is lowered (equivalently, as the inverse temperature grows), the system first undergoes a second order transition to the static phase at temperature $T_\mathrm{static}$. As the temperature is lowered even further, there is a transition into the instanton crystal phase at temperature $T_\mathrm{inst}$, which corresponds to the intersection of the two free energy curves on the plots (marked by black dots).
Since the slopes of the free energy curves at the intersection point are different, this transition is accompanied by a jump in the first derivative of the free energy, henceforth it is of the first order.
On the contrary, for the case $U_2/U_0 = 4.0$ which correspond to the plot Fig.~\ref{temperature} (d), we observe that the system undergoes transition to the instanton crystal phase whithout ever entering the static phase.
Overall, the picture observed in Fig.~\ref{temperature} suggests that as the value of $U_2/U_0$ grows larger, $T_\mathrm{inst}$ moves closer to $T_\mathrm{static}$ until the point where they coincide. For larger values of $U_2/U_0$, $T_\mathrm{inst} > T_\mathrm{static}$ and the transition to the instanton crystal phase happens without the intermediate static phase.

The instanton crystal phase has a rather peculiar feature. As the temperature is lowered, more and more instanton-antiinstanton pairs can fit onto the interval $[0,\beta]$. As a result, there is a series of the first order transitions characterized by the change in the number of the periods of the instanton lattice $m$ by 1. In Fig.~\ref{temperature}, these transitions are marked by red dots.

Finally, we would like to discuss the order of the transition from the normal metal to the instanton crystal phase in the case where is no intermediate static phase involved. This transition is of the second order, which can be understood from the following argument. Let us consider the configurations with a single instanton-antiinstanton pair.
Let us also assume for a moment that there is no nonlocal repulsion, so that the instaton-antiinstanton configuration is described by equations~(\ref{b20}-\ref{n_ktilde}).
The instanton-antiinstanton configuration has a minimal period $W_0$ which is finite. This period corresponds to the solution of the system (\ref{period_equation}-\ref{n_ktilde}) with $k = 0$: $W_0 = 4K(0)/\gamma_{k=0}$. The instanton-antiinstanton configuration has the form $b(\tau) = \gamma k \sn{(\gamma\tau|k)}$, thus, as the period gets close to $W_0$, $k$ gets close to zero and the amplitude of the configuration vanishes in the limit $W\rightarrow W_0+0$. Now, if we take into account the non-local repulsion term, its effects should vanish together with the amplitude of the instanton-antiinstanton configuration.
From this we can conclude that the minimal period of the configuration would stay the same, and the amplitude of the instanton-antiinstanton configuration would vanish in the limit $W\rightarrow W_0+0$ as before. The transition to the instanton crystal phase happens when $1/T = W_0$. As the order parameter vanishes at the transition, we expect it to be of the second order.

%
%

\section{Possible physical origin of the model.\label{origin}}

\subsection{Previous models.}

The idea of investigating the present model, Eqs. (\ref{a2}-\ref{a8}),
originates from the previous studies of superconducting cuprates using the
so-called spin-fermion (SF) model. This phenomenological model has been
proposed in order to enable analytical study of low energy physics of
cuprates~\cite{abanov2003,metlitski2010,efetov2013,wang2014}. The philosophy
underlying this approach is based on integrating out the high energy degrees of
freedom (of the order of the bandwidth) and writing an effective model containing only the
low energy excitations. Of course, after such an integration one obtains a
very complicated effective Lagrangian that can hardly be treated analytically.
In this situation, the only thing that can be done is to simplify the resulting effective
model by reducing it to a form containing a small number of different types of
the excitations. It is important to have a sufficienly simple form of these
excitations and of their interactions.

Originally, Spin Fermion model \cite{abanov2003} was introduced as an effective model containing the 
fermions in the vicinity of the Fermi surface interacting with bosonic
antiferromagnetic waves propagating with vector $\mathbf{Q}$ close to the
antiferromagnetic vector $\mathbf{Q}_{AF}$. The latter are assumed to be the
remnants of the parent insulating AF state. A weak interaction between the fermions and the antiferromagnetic waves is most efficient at $8$ points of the Fermi surface that
can be connected by the vector $\mathbf{Q}_{AF}$ (hot spots). The resulting
interaction is strongly peaked at the wavevector $\mathbf{Q}%
_{0}=(\pi ,-\pi )$ corresponding to the antiferromagnetic order with vector $%
\mathbf{Q}_{AF}$ and is described by a propagator
\begin{equation}
D_{\mathrm{0}}\left( \omega ,\mathbf{q}\right) =\left( \omega
^{2}/v_{s}^{2}+\left( \mathbf{q-Q}_{0}\right) ^{2}+\xi _{AF}^{-2}\right)
^{-1}.  \label{k2}
\end{equation}

In Eq. (\ref{k2}), $v_{s}$ is the spin velocity and $\xi $ is the correlation
length which is supposed to diverge at the antiferromagnetic transtion. It
is important to note that the fermions and bosonic spin waves actually have the same
origin. The spin waves in the effective model are some
complicated collective spin excitations of the bare interacting holes constituting the original
microscopic model (we consider the hole-doped cuprates). At the same time, many details of the microscopic model
are not so important for the 
investigation of universal phenomena such as phase
transitions, symmetry of the phases, etc.

\begin{figure}[t]
 \includegraphics[width=2.57in]{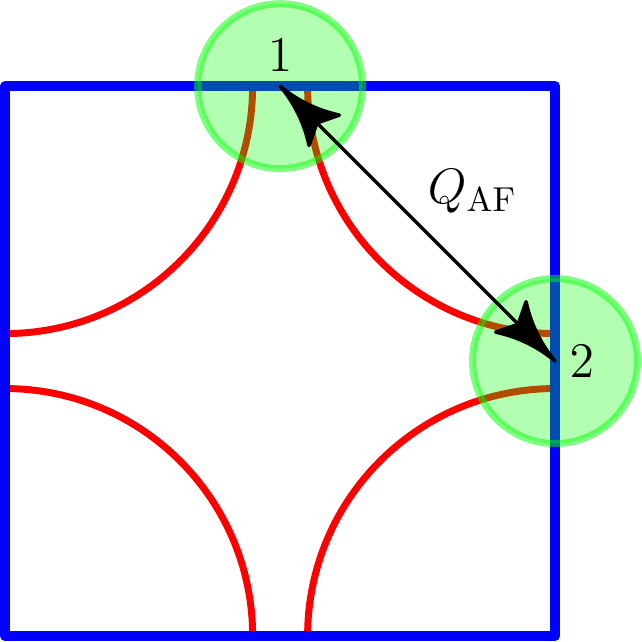}
 \caption{Fermi surface (red) and overlapping hot spots (green).}
 \label{fermi_surface}
\end{figure}

Having integrated out the high energy fermions, one loses the detailed information
about the structure of the lattice.
At the same time, one can use Fermi-liquid-like arguments to conclude that one can still use the basic shape of the Fermi surface (see Fig.~\ref{fermi_surface}).
So, identifying the vector $\mathbf{Q}_{0}$ with the
antiferromagnetic vector $\mathbf{Q}_{AF},$ one can play with the model with
$8$ hotspots and obtain many interesting resuls at low energies.
On the other hand, one can not really exclude the possibility of additional low-lying collective excitations since the Spin Fermion model was introduced on phenomenological basis in the first place.

The use of the propagator for the fermion-fermion interaction, Eq. (\ref{k2}%
), is motivated by the proximity to AF quantum critical point (QCP) where $%
\xi _{AF}\rightarrow \infty $. Then, one can consider only small $\delta
\mathbf{p}\sim 1/\xi _{AF}$ vicinities of the `hot spots' as strongly
affected by the interaction. However, at temperatures relevant for e.g. the
pseudogap state \cite{timusk1999,norman2005,hashimoto2014} this argument
does not have to hold because the experimentally reported correlation lengths
\cite{haug2010,chan2016} are indeed rather small. Moreover, ARPES
experiments \cite{hashimoto2014} show that the effects of the pseudogap
extend well beyond the `hot spots' to the Brillouin zone edges $(\pi
,0),(0,\pi )$ without being significantly weakened.

A different version of the SF model (spin-fermion model with overlapping hot
spots (SFMOHS)) has been introduced in Refs. \cite{volkov1,volkov2,volkov3}.
As $\xi _{AF}$ becomes smaller, the `hot spots' expand and can eventually
overlap and merge forming two `hot regions' (see Fig.~\ref{fermi_surface}). For the latter to
occur the fermionic dispersion in the antinodal region should be shallow,
which is supported by the experimental data \cite{hashimoto2010,kaminski2006}%
. This (SFMOHS) model differs from previously used spin-fermion models with $%
8$ hot spots \cite{abanov2003,metlitski2010,efetov2013,wang2014,pepin2014}
by the\textbf{\ }assumption that the hot spots on the Fermi surface are not
isolated, may overlap and form antinodal `hot regions'. This can happen when
the fermion energies are not far away from the van Hove singularities in the
spectrum of the cuprates, which corresponds to the results of ARPES study \cite%
{hashimoto2010,he,kaminski2006,anzai}.

The two hot regions $1$ and $2$ are centered at the middles of the edges of
the Brillouin zone and can be connected by the vector $\mathbf{Q}_{0}$.
Then, one comes to a description in terms of the fermions located in two bands
with the interaction between the bands. Due to proximity to the van Hove
singularity one can write the spectra of the fermions in the bands near
points $(\pi ,0)$ and $(0,\pi )$ as
\begin{equation}
\varepsilon _{1}(\mathbf{p})=\alpha p_{x}^{2}-\beta p_{y}^{2}-\mu
,\;\varepsilon _{2}(\mathbf{p})=\alpha p_{y}^{2}-\beta p_{x}^{2}-\mu ,
\label{Mod:disp}
\end{equation}%
The momenta $\mathbf{p}$ are counted now from the points $(\pi ,0)$ and $%
(0,\pi )$. In Eq. (\ref{Mod:disp}), $\mu $ is the chemical potential and $%
\alpha $ and $\beta $ are constants.

This discussion makes clear the origin of the bare part of the Hamiltonian $\hat{H}_{\mathrm{0}%
}$, Eqs. (\ref{a3}-\ref{a4}) and of the action $S_{\mathrm{0}}\left[ \chi ,\chi
^{+}\right] $, Eq. (\ref{adef:fermion}). For investigating interaction of
the fermions via the bosonic spin mode $D_{\mathrm{0}}\left( \omega ,\mathbf{%
q}\right) $, Eq. (\ref{k2}), the field theoretical formulation is more
convenient than the Hamiltonian one. As we are intrested in studying
possible states different from the aniferromagnets, it is convenient to
single out the pairs with assumed strong correlations. For example,
one could consider particle-particle pairs when studying superconductivity, or
particle-hole pairs for studying charge density waves (CDW) and loop
currents \cite{abanov2003,volkov1,volkov2,volkov3,efetov2019}.

Here, as in Refs. \cite{abanov2003,volkov1,volkov2,volkov3,efetov2019}, we
are interested in loop-current channel and charge density channel, both
in the vicinity of the vector $\mathbf{Q}_{0}$.
The channels with the non-trivial spin structure of the pairs correspond to weaker interactions and, thus, are not important.

So, singling out the most interesting pairs, we write the interaction term in the
action in the form%
\begin{equation}
S_{\mathrm{int}}\left[ \chi ,\chi ^{+}\right] \rightarrow S^{\left( \mathrm{%
current}\right) }\left[ \chi ,\chi ^{+}\right] +S^{\left( \mathrm{density}%
\right) }\left[ \chi ,\chi ^{+}\right] .  \label{k3}
\end{equation}%
In Eq. (\ref{k3}),
\begin{eqnarray}
&&S^{\left( \mathrm{current}\right) }\left[ \chi ,\chi ^{+}\right] =-\frac{%
3\lambda ^{2}}{8}\int D_{\mathrm{0}}\left( X-X^{\prime }\right)  \notag \\
&&\times \left( \chi ^{+}\left( X^{\prime }\right) \Sigma _{2}\chi \left(
X\right) \right) \left( \chi ^{+}\left( X\right) \Sigma _{2}\chi \left(
X^{\prime }\right) \right) dXdX^{\prime },  \notag \\
&&  \label{k6}
\end{eqnarray}%
stands for attration of the fermionic loop currents, while%
\begin{eqnarray}
&&S^{\left( \mathrm{density}\right) }\left[ \chi ,\chi ^{+}\right] =\frac{%
3\lambda ^{2}}{8}\int D_{\mathrm{0}}\left( X-X^{\prime }\right)  \notag \\
&&\times \left( \chi ^{+}\left( X^{\prime }\right) \Sigma _{1}\chi \left(
X\right) \right) \left( \chi ^{+}\left( X\right) \Sigma _{1}\chi \left(
X^{\prime }\right) \right) dXdX^{\prime }.  \notag \\
&&  \label{k7}
\end{eqnarray}%
stands for the repulsion of the fermion densities oscillating at wave-vector $\mathbf{Q}%
_{0}$ in space, $\lambda $ is the coupling constant of the interaction between the
spin mode the spins of the fermions, and $X=\left( \tau ,\mathbf{r}\right) $
are vectors containing as components the imaginary time and space
coordinates. The fermionic fields $\chi ,\chi ^{+}$
have already been intoduced in Sec. \ref{model}. The signs of the
interactions are unambigously determined by the SF interaction.

The precise momentum dependence of the propagator $D_{\mathrm{0}}\left( \omega ,%
\mathbf{q}\right) $ is not important for us. Therefore, to simplify the model, we replace the propagator by a constant in frequency-momentum space or, equivalently, by $\delta$-function in imaginary time and real space.
Thus, we arrive at Eq.~\eqref{adef:int} for the interaction part of the action.
The inclusion of the long-range part of the Coulomb interaction renormalizes the coupling constant in $S^{(\mathrm{density})}[\xi, \xi^+]$. As a result, we keep the couplings $\tilde{U}_0$ and $U_0$ in Eq.~\eqref{adef:int} as independent constants.


The model only with the interaction $S^{\left( \mathrm{current}\right) }\left[
\chi ,\chi ^{+}\right] $ was studied in Ref.~\cite{volkov3}, and proposed earlier DDW state~\cite{chakravarty}
was obtained in the mean-field approximation.
This state corresponds to the static loop current modulated with the vector $%
\mathbf{Q}_{AF}$ and flowing around the $CO_{2}$ elementary cells.
In this case, the instanton-antiinstanton solutions for the order parameter could be obtained, however,
the free energies of these configurations were higher than that of DDW state.

In Ref.~\cite{efetov2019}, it was argued that a stable instanton-antiinstanton crystal state can be obtained if one adds an extra interaction term $S^{\left( \mathrm{density}%
\right) }\left[ \chi ,\chi ^{+}\right] $ to the model.
However, this result was based on an analytical computation invlolving perturbation expansion up to the second order in the abscence of a ``small parameter''.
In order to get a conclusive proof of the stablity of the instanton-antiinstanton crystal state, we designed the numerical scheme described in Section~\ref{numerics}.
Unfortunately, although the inclusion of $S^{\left( \mathrm{density}%
\right) }\left[ \chi ,\chi ^{+}\right] $ term lowered the free energy of the instanton-antiinstanton configurations,
numerical analysis revealed that DDW state was still energetically more favourable.



On the other hand,
in sections \ref{analytics} and \ref{numerical_analysis}, we have seen that introducing
interaction of the fermions with an additional bosonic mode can drastically
change the situation and make the state with the imaginary-time-dependent
order parameter thermodynamically stable.

\subsection{Interaction of fermions with the bosonic current-like mode.}

Integrating out the high energy modes in microscopic models leaves a lot of
possibilities for low energy modes obtained after this procedure. This is
especially true for cuprates which have a very complicated structure.
The only possibility to model the emerging low energy modes is to introduce them
phenomenologically. 

According to this philosophy, the integration of the high energy degrees of freedom in Spin Fermion model is assumed to lead to the appearance of the bosonic spin modes which are identified with the surviving low-energy antiferromagnetic fluctuations near the antiferromagnetic quantum critical point. In
particular, large values of the propagator $D_{\mathrm{0}}\left( \omega ,%
\mathbf{q}\right) $ at the antiferromagnetic vector $\mathbf{Q}_{AF}$ follow
from this fact.
These bosonic modes couple to the spins of the fermions corresponding to the magnetic origin of the interaction.
Given the phenomenological character of the Spin Fermion model, it is plausible to assume the possibility of the appearance of additional low-energy bosonic modes which couple instead to the magnetic moments induced by the fermion loop currents modulated in space at the wave-vector $\mathbf{Q}_{AF}$.

The bosonic modes described by the Hamiltonian~\eqref{a4b} are precisely of this type.
There, we have a system of oscillators with $\mathcal{\hat{P}}_{\mathbf{q}}$ and coordinates $%
\mathcal{\hat{Q}}_{\mathbf{q}}$ labeled by different wave-vectors $\mathbf{q}$.
It is assumed that these wave-vectors are counted from $Q_\mathrm{AF}$, and their lengths are not large.
The interaction of these modes with the fermionic currents is included in a gauge-invariant manner by coupling the former
to the vector potential $\mathbf{A}_{\mathbf{q}}$ created by fermions (see Eq.~\eqref{a8}).




In the Lagrangian formulation, one writes the action $S_{\mathrm{B}}$, Eq. (%
\ref{adef:boson}), with the coordinate field $a_{\mathbf{q}}\left( \tau
\right) $, while the interaction of the bosonic and fermionic currents is
given by the term $S_{\mathrm{FB}}\left[ \chi ,\chi ^{+},a\right] ,$ Eq. (%
\ref{adef:fb}).

It is worth emphasizing that the Hamiltonian of the model and the corresponding action describe a
system of particles with different $\mathbf{p}$ and $\mathbf{q}$ and, in this
sense, do not differ from the standard many body models. The interesting effects
show up when one studies the condenstate of particle-hole pairs with dominating
contribution at $\mathbf{q}=0$ (see Eq.~(\ref{b2})).

\section{Discussion and Outlook. \label{outlook}}

We have proposed a new thermodynamic model of interacting fermions and
bosonic current-like modes. It does not contain any special features like
long-range or infinite-range interaction.
Using the methods of the field theory we introduced collective boson degrees of freedom and integrated out the fermionic ones, reducing the model to a system of interacting bosons.
The model cannot be solved exactly and, as usual, one
starts with developing mean-field approximation. In our formulation, the
mean-field equations are just equations for the minimum of the bosonic
action. This fact considerably simplifies both analytical and numerical
study.

We have demonstrated that the system can be (in addition to normal metal phase)
either in the stationary phase with a conventional imaginary-time-independent order parameter or in the
instanton crystal phase. The numerical investigation at zero temperature, performed in
Section~\ref{numerical_analysis}, reveals the existence of a quantum phase transition between
these phases. It is also shown that the derivative of the free energy $\partial F/\partial \tilde{\omega}_0$ experiences a jump at the transition indicating that it is a first order transition.

In addition to that, in Section~\ref{numerical_analysis}, we also performed the numerical investigation of the temperature dependence of the free energy of the model.
The results of our calculations indicate that, as the temperature is lowered, the transition from normal metal phase to the instanton crystal phase can happen either via an intermediate stationary phase or directly.
In the former case, the transition from the normal metal phase to the stationary phase is of the second order, while the subsequent transition into the instanton crystal phase is of the first order. In the latter case, the direct transition into the instanton crystal phase is of the second order. As the temperature is lowered further, there is a series of first order transitions corresponding to the change of the number of the periods of the instanton lattice.


As the results have been obtained using the mean-field scheme, it is
important to understand how fluctuations near the saddle-point solution
affect the results. For this purpose, one should make expansion of the
effective action up to the second order and check the eigenvalues of the corresponding
quadratic form.
The stability of the long-range order is endangered by the fluctuations associated with the gapless zero modes,
and in our case, there is a zero mode originating from the translational invariance of the instanton lattice.
On the other hand, we consider the model with at least two spatial dimensions, and the order parameter in the abscence of instantons corresponds to discrete $\mathbb{Z}_2$ symmetry breaking.
Also, the imaginary time acts as an extra dimension, which helps to reduce the effect of the fluctuations.
Overall, the role of the fluctuations is an open question which we plan investigating on in the future.


It is hard to speak about the possible experimental observation of the instanton crystal phase at this stage.
In this paper, we considered only the equilibrium properties of the system.
As a result, we can only suggest to look for the discontinuity in the derivative of the free energy.
The significant amount of information about the system, on the other hand, can be obtained by measuring its response to various probes.
As a consequence, another important direction of our future studies is the analysis of the real-time correlation functions. This task is challenging and deserves a special treatment. Therefore, we decided not to touch the subject in this paper.



\begin{acknowledgments}
Financial support of Deutsche Forschungsgemeinschaft (Projekt~EF~11/10-1)
and of the Ministry of Science and Higher Education of the Russian
Federation in the framework of Increase Competitiveness Program of NUST
\textquotedblleft MISiS\textquotedblright (Nr.~K2-2017-085") is greatly
appreciated.
\end{acknowledgments}

\appendix

\section{Inversion of the integral operator describing the effective fermion interaction\label%
{inversion}}

The interaction kernel $K(\tau-\tau^\prime|\omega_\mathbf{q})$ was defined
in Eq.~\eqref{kernel:def} as
\begin{multline}
K(\tau - \tau^\prime|\omega_\mathbf{q}) = (U_0 + U_2)\delta(\tau -
\tau^\prime) - \\
\* - U_2\cdot K_0(\tau - \tau^\prime|\omega_\mathbf{q}),
\end{multline}
where the function
\begin{equation}
K_0(\tau - \tau^\prime|\omega_\mathbf{q}) = \cfrac{\omega_\mathbf{q}\cosh{%
\left[\omega_\mathbf{q}\left(\frac{\beta}{2} -
|\tau-\tau^\prime|\right)\right]}}{2\sinh{\frac{\omega_0}{2}}}
\end{equation}
is the greens function for a certain differential operator (see Eqs.~%
\eqref{green0} and~\eqref{green1})
\begin{equation}
\left[-\cfrac{1}{\omega_\mathbf{q}^2}\left(\cfrac{d}{d\tau}\right)^2 + 1%
\right] K_0(\tau-\tau^\prime|\omega_\mathbf{q}) = \delta(\tau- \tau^\prime)
\label{green}
\end{equation}

As it turns out, that is all we need to construct the inverse of the
integral operator described by kernel $K(\tau-\tau^\prime)$. Let us denote
the kernel for the inverse operator as $K^{-1}(\tau-\tau^\prime)$. It should
satisfy
\begin{equation}
\int\limits_0^{\beta} d\tau^\prime K(\tau -
\tau^\prime)K^{-1}(\tau^\prime-\tau^{\prime\prime}) = \delta(\tau -
\tau^{\prime\prime}).  \label{inv_def}
\end{equation}
Let us seek $K^{-1}(\tau-\tau^\prime)$ in the form
\begin{equation}
K^{-1}(\tau-\tau^\prime) = A\cdot(\tau-\tau^\prime) + g(\tau-\tau^\prime),
\label{inv_form}
\end{equation}
where $A$ is a yet unknown parameter and $g(\tau-\tau^\prime)$ is an unknown
function. In order to determine them, let us plug the ansatz~\eqref{inv_form}
into Eq.~\eqref{inv_def}.

\begin{multline}
\delta(\tau-\tau^{\prime\prime}) =
A(U_0+U_2)\delta(\tau-\tau^{\prime\prime}) - AU_2\cdot
K_0(\tau-\tau^{\prime\prime}) + \\
\* + (U_0+U_2)g(\tau-\tau^{\prime\prime}) - U_2\cdot\int\limits_0^{\beta}
d\tau^{\prime} K_0(\tau-\tau^{\prime})g(\tau^{\prime}-\tau^{\prime\prime})
\end{multline}
It seems reasonable to put $A = (U_0+U_2)^{-1}$. Thus, we obtain the
following integral equation for the unknown function $g(\tau-\tau^\prime)$:
\begin{multline}
(U_0+U_2)g(\tau-\tau^{\prime\prime}) - U_2\cdot\int\limits_0^{\beta}
d\tau^{\prime} K_0(\tau-\tau^{\prime})g(\tau^{\prime}-\tau^{\prime\prime}) =
\\
\* = \cfrac{U_2}{U_0+U_2}K_0(\tau-\tau^{\prime\prime}),
\end{multline}
or, equivalently,
\begin{multline}
g(\tau-\tau^{\prime\prime}) - \cfrac{U_2}{U_0+U_2}\cdot\int\limits_0^{\beta}
d\tau^{\prime} K_0(\tau-\tau^{\prime})g(\tau^{\prime}-\tau^{\prime\prime}) =
\\
\* = \cfrac{U_2}{(U_0+U_2)^2}K_0(\tau-\tau^{\prime\prime}).  \label{int_eq}
\end{multline}
Let us apply the differential operator from Eq.~\eqref{green} to the both
sides of Eq.~\eqref{int_eq}. This way, we get
\begin{multline}
\left[-\cfrac{1}{\omega_\mathbf{q}^2}\left(\cfrac{d}{d\tau}\right)^2 +
\left(1 - \cfrac{U_2}{U_0+U_2}\right)\right] g(\tau-\tau^{\prime\prime}) = \\
\* = \cfrac{U_2}{(U_0+U_2)^2}\delta(\tau- \tau^{\prime\prime}).
\end{multline}
We can further rewrite it as
\begin{multline}
\left[-\cfrac{U_0+U_2}{U_0\omega_\mathbf{q}^2}\left(\cfrac{d}{d\tau}%
\right)^2 + 1\right] g(\tau-\tau^{\prime\prime}) = \\
\* = \cfrac{U_2}{U_0(U_0+U_2)}\delta(\tau- \tau^{\prime\prime}).
\end{multline}
This equation has the same functional form as the equation \eqref{green}
which $K_0(\tau-\tau^\prime|\omega_\mathbf{q})$ satisfies. As a result, we
can write the solution right away:
\begin{equation}
g(\tau-\tau^\prime) = \cfrac{U_2}{U_0(U_0+U_2)}\times K_0(\tau-\tau^\prime|%
\tilde{\omega}_\mathbf{q}).
\end{equation}
where
\begin{equation}
\tilde{\omega}_\mathbf{q} = \sqrt{\frac{U_0}{U_0+U_2}}\times \omega_\mathbf{q%
}.
\end{equation}

\section{Transformation of the fermionic part of the free energy functional.\label{el_transformation}}

Let us focus on the part of the free energy functional~\eqref{fenergy_functional} which originates from integrating out the fermionic degrees of freedom:
\begin{equation}
 \frac{\BF_\mathrm{ferm}[b(\tau)]}{TV} = -2
\int\frac{d\mathbf{p}}{(2\pi)^2} \int\limits_0^\beta d\tau\tr\left[\ln{\check h_{\mathbf{p}}}\right]_{\tau,\tau}.
\end{equation}
Here, we neglect the field $b_1(\tau)$, so that
\begin{equation}
 \check{h}_\mathbf{p}(\tau) = \check{h}_{0\mathbf{p}}(\tau) - b(\tau)\check\Sigma_2,
\end{equation}
where $\check{h}_{0\mathbf{p}}(\tau)$ is defined in Eq.~\eqref{a22a}

We can rewrite equivalently
\begin{equation}
 \int\limits_0^\beta d\tau\tr\left[\ln{\check h_{\mathbf{p}}}\right]_{\tau,\tau} = \tr_{\tau,s} \ln{\left[\check h_{\mathbf{p}}(\tau)\right]} = \ln \det_{\tau,s} \left[\check h_{\mathbf{p}}\right].
\end{equation}
Here, $\tr_{\tau,s}$ stands for combined trace in the subspace of anti-periodic functions and in the subspace of the bands $1$ and $2$, while $\det_{\tau, s}$ stands for the combined determinant in the same subspaces.

In the following, it is also convenient to regularize $\BF_\mathrm{ferm}$ by subtracting the constant term corresponding to the normal metal configuration $b(\tau)\equiv0$. This way, we obtain
\begin{equation}
 \frac{\BF_\mathrm{ferm}[b(\tau)] - \BF_\mathrm{ferm}[0]}{TV} = -2\int \frac{d\mathbf{p}}{(2\pi)^2} \ln\frac{\det_{\tau,s}\left[h_{\mathbf{p}}\right]}{\det_{\tau,s}\left[h_{0\mathbf{p}}\right]}\label{fenergy_el}
\end{equation}




It is hard to work directly with the functional determinants.
However, we can re-express the ratio of two functional determinants in terms of the time-ordered exponentials of the corresponding operators
(It is the standard trick in the field of Determinant Monte-Carlo. For proof of this relation, see, for example, \cite{blankenbecler-81}):
\begin{multline}
 \cfrac{\det_{\tau,s}\left[\check{\mathrm{I}}\partial_\tau + \left(\varepsilon^+_\mathbf{p}\check{\mathrm{I}} + \varepsilon^-_\mathbf{p}\check\Sigma_3 - b(\tau)\check\Sigma_2\right)\right]}{\det_{\tau,s}\left[\check{\mathrm{I}}\partial_\tau + \left(\varepsilon^+_\mathbf{p}\check{\mathrm{I}} + \varepsilon^-_\mathbf{p}\check\Sigma_3\right)\right]}
 = \\ \* =
 \cfrac{\det_s\left[\check{\mathrm{I}} + \BT e^{-\int_0^\beta d\tau (\varepsilon^+_\mathbf{p}\check{\mathrm{I}} +\varepsilon^-_\mathbf{p}\check\Sigma_3 - b(\tau)\check\Sigma_2)}\right]}
 {\det_s\left[\check{\mathrm{I}}+ e^{-\beta(\varepsilon^+_\mathbf{p}\check{\mathrm{I}} +\varepsilon^-_\mathbf{p}\check\Sigma_3)}\right]},\label{gelfand_yaglom}
\end{multline}
where $\BT$ is the time ordering operator and we used the explicit expressions for $\check{h}_\mathbf{p}(\tau)$ and $\check{h}_{0\mathbf{p}}(\tau)$.
We should note that this particular form with $\check{\mathrm{I}} + \BT \exp[\dotsb]$ is attributed to the fact that we considered the functional determinants of the operators with anti-periodic boundary conditions.

The expression in Eq.~\eqref{gelfand_yaglom} can be further simplified.
First, for $2\times2$ matrices, one can show by direct substitution that
\begin{equation}
\det{\left[\check{\mathrm{I}} + \check{A}\right]} = 1 + \det{\check{A}} + \tr{\check{A}}\label{identity2x2}
\end{equation}
Secondly, since $\check{\mathrm{I}}$ commutes with any $2\times2$ matrices, we can write in Eq.~\eqref{gelfand_yaglom}
\begin{multline}
 \BT e^{-\int_0^\beta d\tau (\varepsilon^+_\mathbf{p}\check{\mathrm{I}} +\varepsilon^-_\mathbf{p}\check\Sigma_3 - b(\tau)\check\Sigma_2)}
 = \\ \* =
 e^{-\beta\varepsilon^+_\mathbf{p}} \BT e^{-\int_0^\beta d\tau (\varepsilon^-_\mathbf{p}\check\Sigma_3 - b(\tau)\check\Sigma_2)}\label{com1}
\end{multline}
and
\begin{equation}
 e^{-\beta(\varepsilon^+_\mathbf{p}\check{\mathrm{I}} +\varepsilon^-_\mathbf{p}\check\Sigma_3)} = e^{-\beta\varepsilon^+_\mathbf{p}}e^{-\beta\varepsilon^-_\mathbf{p}\check\Sigma_3}.\label{com2}
\end{equation}
Finally, we should note, that for the time-ordered operator in Eq.~\eqref{com1},
\begin{multline}
 \det_s\left[\BT e^{-\int_0^\beta d\tau (\varepsilon^-_\mathbf{p}\check\Sigma_3 - b(\tau)\check\Sigma_2)}\right]
 = \\ \* =
 \BT e^{-\int_0^\beta d\tau \mathrm{tr}{\left[\varepsilon^-_\mathbf{p}\check\Sigma_3 - b(\tau)\check\Sigma_2\right]}} = 1,\label{liouville}
\end{multline}
which follows from Liouville's theorem and from the fact that Pauli matrices are traceless.

Substituting Eqs.~\eqref{identity2x2}, \eqref{com1}, \eqref{com2} and~\eqref{liouville} alltogether into Eq.~\eqref{gelfand_yaglom}, we obtain
\begin{multline}
 \cfrac{\det_s\left[\check{\mathrm{I}} + \BT e^{-\int_0^\beta d\tau (\varepsilon^+_\mathbf{p}\check{\mathrm{I}} +\varepsilon^-_\mathbf{p}\check\Sigma_3 - b(\tau)\check\Sigma_2)}\right]}
 {\det_s\left[\check{\mathrm{I}}+ e^{-\beta(\varepsilon^+_\mathbf{p}\check{\mathrm{I}} +\varepsilon^-_\mathbf{p}\check\Sigma_3)}\right]}
 = \\ \* =
 \cfrac{2\cosh{\beta\varepsilon_\mathbf{p}^+} + \mathrm{tr}\left[\BT e^{-\int_0^\beta d\tau (\varepsilon^-_\mathbf{p}\check\Sigma_3 - b(\tau)\check\Sigma_2)}\right]}{2(\cosh{\beta\varepsilon_\mathbf{p}^+} + \cosh{\beta\varepsilon_\mathbf{p}^-})}.
\end{multline}
Now, we can rewrite Eq.~\eqref{fenergy_el} as
\begin{multline}
  \frac{\BF_\mathrm{ferm}[b(\tau)] - \BF_\mathrm{ferm}[0]}{TV} = 
 - 
 2\int \cfrac{d\mathbf{p}}{(2\pi)^2}\times \\ \* \times
 \ln{\cfrac{2\cosh{\beta\varepsilon_\mathbf{p}^+} + \mathrm{tr}\left[\BT e^{-\int_0^\beta d\tau (\varepsilon^-_\mathbf{p}\check\Sigma_3 - b(\tau)\check\Sigma_2)}\right]}{2(\cosh{\beta\varepsilon_\mathbf{p}^+} + \cosh{\beta\varepsilon_\mathbf{p}^-})}}\label{fenergy_basis}
\end{multline}

Instead of the normal metal configuration, we could have used a static configuration $b(\tau)\equiv \gamma$ to regularize the fermionic part of the free energy functional. In this case, one can write
\begin{multline}
 \frac{\BF_\mathrm{ferm}[b(\tau)] - \BF_\mathrm{ferm}[\gamma]}{TV} = \frac{\BF_\mathrm{ferm}[b(\tau)] -  \BF_\mathrm{ferm}[0]}{TV} - \\ \* -
 \frac{\BF_\mathrm{ferm}[\gamma] - \BF_\mathrm{ferm}[0]}{TV} 
 =
 - 
 2\int \cfrac{d\mathbf{p}}{(2\pi)^2}\times \\ \* \times
 \ln{\cfrac{2\cosh{\beta\varepsilon_\mathbf{p}^+} + \mathrm{tr}\left[\BT e^{-\int_0^\beta d\tau (\varepsilon^-_\mathbf{p}\check\Sigma_3 - b(\tau)\check\Sigma_2)}\right]}{2\left(\cosh{\beta\varepsilon_\mathbf{p}^+} + \cosh{\beta\sqrt{(\varepsilon_\mathbf{p}^-)^2+\gamma^2}}\right)}}.\label{fenergy_staticreg}
\end{multline}

\section{Correction to the free energy of the instanton-antiinstanton configurations due to the non-local repulsion.\label{correction_integral_calculation}}

Let us consider the integral appearing in Eq.~\eqref{correction_iai},
which describes the correction to the free energy of the instanton-antiinstanton configuration due to repulsive interaction:
\begin{multline}
\Delta F_\mathrm{inst}^\mathrm{repul} = \\ \* = \cfrac{T U_2}{U_0^2}\iint\limits_{0}^{\beta }d\tau d\tau ^{\prime
}K_{0}(\tau -\tau ^{\prime }|\omega _{0})b_{\mathrm{inst}}^{m,k}(\tau )b_{%
\mathrm{inst}}^{m,k}(\tau ^{\prime }),\label{correction_integral}
\end{multline}
Here, $b_{\mathrm{inst}}^{m,k}(\tau) = \gamma k \sn{(\gamma \tau|k)}$ is the configuration with $m$ instanton-antiinstanton pairs and the parameter $\gamma$ is fixed by the choice of the parameters $m$ and $k$ according to
Eqs.~\eqref{period_equation}, \eqref{self_consistency}, \eqref{kappa} and~\eqref{n_ktilde}.%

In order to evaluate the integral, it is convenient to use the known Fourier decomposition of the snoidal Jacobi function (see Ref.~\cite{whittaker}):
\begin{multline}
 b_{\mathrm{inst}}^{m,k}(\tau) = \frac{\gamma\pi}{K(k)}\sum\limits_{n=1}^{+\infty} \frac{\sin{\left[\frac{\gamma\pi(2n-1)}{2K(k)}\tau\right]}}{\sinh{\left[\frac{(2n-1)\pi K(k^\prime)}{2K(k)}\right]}} = \\ \* =
 \frac{\gamma\pi i}{2K(k)} \sum\limits_{n=-\infty}^{+\infty} \frac{\exp{\left[-i 2\pi m(2n-1)T\tau\right]}}{\sinh{\left[\frac{(2n-1)\pi K(k^\prime)}{2K(k)}\right]}},\label{snoidal_fourier}
\end{multline}
where the complementary modulus is ${k^\prime}^2 = 1 - k^2$ and we used the fact that $mT = \gamma/(4K(k))$.
In addition to that, we need the Fourier decomposition for the kernel $K_0(\tau-\tau^\prime | \omega_0)$:
\begin{equation}
 K_0(\tau-\tau^\prime | \omega_0) = \sum\limits_{\Omega_n} \frac{T\omega_0^2}{\omega_0^2 + \Omega_n^2} e^{-i\Omega_n (\tau - \tau^\prime)},\label{kernel_fourier}
\end{equation}
where $\Omega_n = 2\pi T n$ are bosonic Matsubara frequencies.
Subsituting Eqs.~\eqref{snoidal_fourier} and~\eqref{kernel_fourier} into Eq.~\eqref{correction_integral},
one obtains
\begin{multline}
 \Delta F_\mathrm{inst}^\mathrm{repul} = \frac{T^2 U_2}{U_0^2} \sum\limits_{\Omega_l} K_{0\Omega_l} b_{\Omega_l} b_{-\Omega_l}
 = \\ \* =
 \frac{U_2}{U_0^2}\frac{4\gamma^2}{\pi^2} \sum\limits_{n = -\infty}^{+\infty} \frac{1}{1 + \left[(2n-1)\frac{\pi\gamma}{2\omega_0 K(k)}\right]^2}
 \times \\ \* \times
 \frac{1}{\left(\frac{4K(k)}{\pi^2}\sinh{\left[\frac{(2n-1)\pi K(k^\prime)}{2K(k)}\right]}\right)^2}.\label{correction_series}
\end{multline}

In the limit $k\rightarrow 1-0$, we can replace $\gamma$ by the corresponding value $\gamma_0 = \gamma_{T=0}$ for the static configuration.
Also, in this limit $K(k^\prime) = K(0) = \pi/2$.
In Eq.~\eqref{correction_series}, the first factor is naturally cut off at $|n|\sim K(k)\omega_0/\gamma_0$.
At the same time, the nonlinearity of hyperbolic sine kicks in for $|n|\sim K(k)$.
If we assume that $\omega_0/\gamma_0 \ll 1$, the convergence of the series is then determined by the first factor,
thus one can safely replace the hyperbolic sine by its argument.
As a result, Eq.~\eqref{correction_series} transforms into
\begin{multline}
 \Delta F_\mathrm{inst}^\mathrm{repul} \approx \frac{U_2}{U_0^2}\frac{4\gamma_0^2}{\pi^2}
 \times \\ \* \times
 \sum\limits_{n = -\infty}^{+\infty} \frac{1}{1 + \left[(2n-1)\frac{\pi\gamma_0}{2\omega_0 K(k)}\right]^2}\frac{1}{(2n-1)^2}.
 \label{correction_series_approx}
\end{multline}
The series appearing in this equation can be summed in the closed form to give
\begin{equation}
 \Delta F_\mathrm{inst}^\mathrm{repul} = \frac{U_2 \gamma_0^2}{U_0^2}\left[1 - \frac{\gamma_0}{\omega_0 K(k)}\tanh{\frac{\omega_0K(k)}{\gamma_0}}\right].
\end{equation}
Finally, in the limit $\omega/\gamma_0 \ll 1/K(k)$, we can taylor expand hyperbolic tangent up to the third order to obtain
\begin{equation}
 \Delta F_\mathrm{inst}^\mathrm{repul} = \frac{U_2 \gamma_0^2}{U_0^2}\times \frac{1}{3}\left(\frac{\omega_0 K(k)}{\gamma_0}\right)^2.
\end{equation}

\section{Details of the implementation of the numerical scheme.}

In this Appendix, we would like to mention several details which are important for the speed and stability of the implementation.

The most crucial part is the calculation of the time-ordered exponential $\check{U}_\mathbf{p}[b_i]$.
In the case of large period of the configuration $W$, it is easy to overflow the exponent of the floating-point numbers used to store the matrix elements. This problem may be overcome if $\check{U}_\mathbf{p}[b_i]$ is calculated in extended-precision arithmetics.
A complementary solution to this problem is to use the different regularization of the fermionic part of the free energy functional. In Eq.~\eqref{fenergy}, we subtracted the constant corresponding to the fermionic part of the free energy of the normal metal configuration. Instead, we could subtract $\BF_\mathrm{ferm}[\gamma_T]$ where $\gamma_T$ is the solution of the static gap equation~\eqref{self_consistency:static} with $U_0$ replaced by $U_0+U_2$. (The idea here is not to put $U_2 = 0$ but to neglect the non-local quadratic part of the free energy functional.)
In this case, equations~\eqref{discrete_f} and~\eqref{grad_f} can be rearranged in such a manner, that, instead of $\check{U}_\mathbf{p}[b_i]$, one needs to calculate
\begin{multline}
 e^{-W\sqrt{(\varepsilon_\mathbf{p}^-)^2 + \gamma_T^2}}\check{U}_\mathbf{p}[b_i] = \\ \* = 
 \prod\limits_{i = N}^1 e^{-\Delta\tau\sqrt{(\varepsilon_\mathbf{p}^-)^2 + \gamma_T^2}} e^{-\Delta\tau\left(\varepsilon_\mathbf{p}^-\check\Sigma_3 - b_i \check\Sigma_2\right)},
\end{multline}
which happens to be much more numerically stable.

The calculation of non-local part of free energy and its gradient, Eqs.~\eqref{discrete_quad_nloc} and~\eqref{grad_quad_nloc}
requires the computation of matrix-vector products
\begin{equation}
 \sum\limits_{j = 1}^{N} \tilde K_{i-j} b_j,
\end{equation}
where the matrix $\tilde K_{i-j}$ is
\begin{equation}
 \tilde{K}_{i-j} = \tilde K_0(\tau_i-\tau_j|\tilde\omega_0).
\end{equation}
In fourier space, matrix-vector product of this type reduces to the element-wise multiplication of vectors.
As a result, these products can be efficiently computed via the following sequence of steps: calculate fast fourier transform of $b_j$; multiply it element-wise by the precomputed fourier transform of $\tilde K_{i,j}$; make another fast fourier transform.

For the numerical calculations in this paper, we implemented the numerical scheme using the programming language \textsl{Julia}~\cite{Julia-17}.
This language combines the fast prototyping of \textsl{Python}, \textsl{Matlab} and \textsl{Mathematica} with the speed of \textsl{Fortran}, \textsl{C} and \textsl{C++}.
For optimization, we used L-BFGS algorithm~\cite{Liu-89,Nocedal-06} implemented in \textsl{Optim.jl} library~\cite{Optim-18}.




\FloatBarrier


%

\clearpage
\onecolumngrid

\begin{center}
  \bf\large
  Supplementary Material\\ for the paper\\ ``Phase transition into Instanton Crystal.''
\end{center}\bigskip
  
\twocolumngrid
  
\setcounter{figure}{0}
\setcounter{equation}{0}
\setcounter{table}{0}
\setcounter{section}{0}
  
\renewcommand{\thefigure}{S\arabic{figure}}

\renewcommand{\theequation}{S\arabic{equation}}

\renewcommand{\thetable}{S\arabic{table}}

\renewcommand{\thesection}{S\Roman{section}}

All equation numbers, figure numbers and reference numbers without prefix ``S'' refer to the respective numbers in the main text.

\section{Instanton-Antiinstanton solutions of Eq.~(4.1).\label{jacobi_solutions}}

\subsection{Mean-field equations}

In Appendix~\ref{el_transformation}, we showed how the electronic part of the free energy can be transformed into the form suitable both for analytical and numerical investigations.
With the help of this transformation, we can rewrite Eq.~\eqref{fenergy_functional} in the case of $\tilde{U}_0=0$ and $U_2 =0$ as
\begin{multline}
  \cfrac{\BF[b(\tau)] - \BF[0]}{TV} = \cfrac{1}{{U_0}}\int\limits_0^\beta d\tau b^2(\tau)
 - 
 2\int \cfrac{d\mathbf{p}}{(2\pi)^2}\times \\ \* \times
 \ln{\cfrac{2\cosh{\beta\varepsilon_\mathbf{p}^+} + \mathrm{tr}\left[\BT e^{-\int_0^\beta d\tau (\varepsilon^-_\mathbf{p}\check\Sigma_3 - b(\tau)\check\Sigma_2)}\right]}{2(\cosh{\beta\varepsilon_\mathbf{p}^+} + \cosh{\beta\varepsilon_\mathbf{p}^-})}}\label{fenergy_basis2}
\end{multline}
Let us define
\begin{equation}
 \check{U}_\mathbf{p}(\tau_2, \tau_1) = \BT e^{-\int_{\tau_1}^{\tau_2} d\tau (\varepsilon^-_\mathbf{p}\check\Sigma_3 - b(\tau)\check\Sigma_2)}.\label{evolution:op}
\end{equation}
Putting the first variation of Eq.~\eqref{fenergy_basis2} to zero, we obtain
\begin{multline}
 \cfrac{ b(\tau)}{{U_0}} = \int \cfrac{d\mathbf{p}}{(2\pi)^2} \cfrac{\mathrm{tr}\left[\check{U}_\mathbf{p}(\beta,\tau)\check\Sigma_2\check{U}_\mathbf{p}(\tau,0)\right]}{2\cosh{\beta\varepsilon_\mathbf{p}^+} + \mathrm{tr}\left[\check{U}_\mathbf{p}(\beta,0)\right]}
 = \\ \* =
 \int \cfrac{d\mathbf{p}}{(2\pi)^2} \cfrac{\mathrm{tr}\left[\check\Sigma_2\check{U}_\mathbf{p}(\tau,0)\check{U}_\mathbf{p}(\beta,\tau)\right]}{2\cosh{\beta\varepsilon_\mathbf{p}^+} + \mathrm{tr}\left[\check{U}_\mathbf{p}(\beta,0)\right]}\label{lgap:time_ordered}
\end{multline}
Comparing the right-hand sides of Eqs.~\eqref{lgap:time_ordered} and~\eqref{lgap:local},
one can argue that the fermion greens function at coinciding times is
\begin{equation}
 \check{G}_\mathbf{p}(\tau,\tau) = -\cfrac{\check{U}_\mathbf{p}(\tau,0)\check{U}_\mathbf{p}(\beta,\tau)}{2\cosh{\beta\varepsilon_\mathbf{p}^+} + \mathrm{tr}\left[\check{U}_\mathbf{p}(\beta,0)\right]}.\label{green:equal_time}
\end{equation}
(One can write the full expression for $\check{G}_\mathbf{p}(\tau,\tau^\prime)$ in terms of operators $\check{U}_\mathbf{p}(\tau_1,\tau_2)$ and then put $\tau=\tau^\prime$ to obtain this equation. The full expression can be found in \cite{blankenbecler-81}.)

As we have already pointed out, the determinant of $\check{U}_\mathbf{p}(\beta,0)$ should be equal to 1.
Thus, the two eigenvalues of $\check{U}_\mathbf{p}(\beta,0)$ must be $\lambda$ and $1/\lambda$.
We shall introduce the parameter $\kappa_\mathbf{p}$ so that $\lambda = e^{\beta\kappa_\mathbf{p}}$.
With this definition, we can write
\begin{equation}
 \mathrm{tr}\left[\check{U}_\mathbf{p}(\beta,0)\right] = \lambda + 1/\lambda = 2\cosh{\beta\kappa_\mathbf{p}}\label{trace_form}
\end{equation}

\subsection{Solution of the gap equation~\eqref{lgap:time_ordered} in terms of Jacobi elliptic functions}

Let us introduce
\begin{subequations}\label{pseudospins}
\begin{align}
 X_\mathbf{p}(\tau) & = -\mathrm{tr}\left[\check\Sigma_1\check{G}_\mathbf{p}(\tau,\tau)\right],\\
 Y_\mathbf{p}(\tau) & = -\mathrm{tr}\left[\check\Sigma_2\check{G}_\mathbf{p}(\tau,\tau)\right],\\
 Z_\mathbf{p}(\tau) & = -\mathrm{tr}\left[\check\Sigma_3\check{G}_\mathbf{p}(\tau,\tau)\right].
\end{align}
\end{subequations}
With the help of this definitions, we can write
\begin{multline}
 2\check{G}_\mathbf{p}(\tau,\tau) = \mathrm{tr}[\check{G}_\mathbf{p}(\tau,\tau)]\check{\mathrm{I}} - \\ \* - X_\mathbf{p}(\tau)\check\Sigma_1 - Y_\mathbf{p}(\tau)\check\Sigma_2 - Z_\mathbf{p}(\tau)\check\Sigma_3.\label{matrix_decomposition}
\end{multline}

At the same time,
\begin{align}
 \partial_\tau \check{G}_\mathbf{p}(\tau,\tau) & = -\left[\varepsilon_\mathbf{p}^-\check\Sigma_3 - b(\tau)\check\Sigma_2, \check{G}_\mathbf{p}(\tau,\tau)\right], \label{gequation}
\end{align}
where we used the definition~\eqref{evolution:op} of $\check{U}_\mathbf{p}(\tau_2,\tau_1)$ as time-ordered exponentials.
Substituting Eq.~\eqref{matrix_decomposition} into Eq.~\eqref{gequation} and using Eqs.~\eqref{pseudospins}, we obtain the following system of equations for functions $X_\mathbf{p}(\tau)$, $Y_\mathbf{p}(\tau)$ and $Z_\mathbf{p}(\tau)$:
\begin{subequations}\label{pseudospins:evolution}
 \begin{align}
  \dot{X}_\mathbf{p}(\tau) & = 2ib(\tau) Z_\mathbf{p}(\tau) + 2i\varepsilon_\mathbf{p}^- Y_\mathbf{p}(\tau),\label{pseudospins:x_evol}\\
  \dot{Y}_\mathbf{p}(\tau) & = -2i\varepsilon_\mathbf{p}^- X_\mathbf{p}(\tau),\label{pseudospins:y_evol}\\
  \dot{Z}_\mathbf{p}(\tau) & = -2ib(\tau) X_\mathbf{p}(\tau)\label{pseudospins:z_evol}.
 \end{align}
\end{subequations}
In addition to this, we have the gap equation~\eqref{lgap:time_ordered}, which we can write as
\begin{equation}
 \cfrac{b(\tau)}{U_0+U_2} = \int\cfrac{d\mathbf{p}}{(2\pi)^2} Y_\mathbf{p}(\tau).\label{lgap:pseudospins}
\end{equation}

We shall note, that we can satisfy Eq.~\eqref{lgap:pseudospins} if we assume, that the functions $Y_\mathbf{p}(\tau)$ are proportional to $b(\tau)$ for all $\mathbf{p}$:
\begin{equation}
 Y_\mathbf{p}(\tau) = A_\mathbf{p} b(\tau).\label{ansatz:y}
\end{equation}
If we subsitute this ansatz into Eq.~\eqref{pseudospins:y_evol}, we fix the function $X_\mathbf{p}(\tau)$:
\begin{equation}
 X_\mathbf{p}(\tau) = \cfrac{i}{2\varepsilon_\mathbf{p}^-} \dot{Y}_\mathbf{p}(\tau) = \cfrac{iA_\mathbf{p}}{2\varepsilon_\mathbf{p}^-}\dot{b}_y(\tau).\label{ansatz:x}
\end{equation}
Anologously, we obtain from Eq.~\eqref{pseudospins:z_evol} that
\begin{equation}
 \dot{Z}_\mathbf{p}(\tau) = \cfrac{A_\mathbf{p}}{2\varepsilon_\mathbf{p}^-}\cfrac{d}{d\tau}(b^2(\tau)).
\end{equation}
Thus, up to some constant term $B_\mathbf{p}$
\begin{equation}
 Z_\mathbf{p}(\tau) = \cfrac{A_\mathbf{p}}{2\varepsilon_\mathbf{p}^-} b^2(\tau) + B_\mathbf{p}.\label{ansatz:z}
\end{equation}
Finally, we can subsitute Eqs.~\eqref{ansatz:y}, \eqref{ansatz:x} and~\eqref{ansatz:y} into Eq.~\eqref{pseudospins:x_evol} thus obtaining the equation determining the stationary field $b(\tau)$:
\begin{equation}
 \ddot{b}_y(\tau) - \left(4(\varepsilon_\mathbf{p}^-)^2 + 4\varepsilon_\mathbf{p}^-\cfrac{B_\mathbf{p}}{A_\mathbf{p}}\right) b(\tau) - 2b^3(\tau) = 0.\label{beq1}
\end{equation}
Or, equivalently, if we multiple it by $b(\tau)$ and integrate,
\begin{equation}
  (\dot{b}_y(\tau))^2 - \left(4(\varepsilon_\mathbf{p}^-)^2 + 4\varepsilon_\mathbf{p}^-\cfrac{B_\mathbf{p}}{A_\mathbf{p}}\right) b^2(\tau) - b^4(\tau) = const.\label{beq2}
\end{equation}
Since Eqs.~\eqref{beq1} and~\eqref{beq2} are supposed to be valid for any value of $\mathbf{p}$, the $\mathbf{p}$-dependent terms should actually be constant:
\begin{equation}
 4(\varepsilon_\mathbf{p}^-)^2 + 4\varepsilon_\mathbf{p}^-\cfrac{B_\mathbf{p}}{A_\mathbf{p}} = -\Delta.\label{constant_fixation}
\end{equation}
Equation~\eqref{beq2} can be interpreted as the law of energy conservation for a particle in a potential $V(b) = \Delta b^2 - b^4 +const$.
As a consequence, if we want Eq.~\eqref{beq2} to have non-trivial periodic solution, the potential should have a local-minima, in the vicinity of which the particle can oscillate. Thus, we shall assume that $\Delta > 0$.

Let us make the substitution $b(\tau) = \zeta w(\gamma\tau)$ in Eq.~\eqref{beq1}.
This way the equation tranforms into
\begin{equation}
 \ddot{w}(\tau) + \cfrac{\Delta}{\gamma^2}w(\tau) - 2 \left(\cfrac{\zeta}{\gamma}\right)^2 w^3(\tau) = 0.\label{eq:jacobi_temp}
\end{equation}
Let us choose $\zeta$ and $\gamma$ in such a way, that $\zeta/\gamma = k \in [0,1]$ and $\gamma = \sqrt{\Delta/(1+k^2)}$.
Then, Eq.~\eqref{eq:jacobi_temp} appears as
\begin{equation}
 \ddot{w}(\tau) + (1+k^2)w(\tau) - 2k^2 w^3(\tau) = 0.\label{eq:jacobi}
\end{equation}
The solution to this equation is Jacobi elliptic function $sn(\tau|k)$.
Henceforth, we have shown that the gap equation~\eqref{lgap:time_ordered} has non-trivial solutions of the form
\begin{equation}
 b(\tau) = \gamma k \times sn(\gamma\tau|k).\label{b_form}
\end{equation}

Finally, we should derive the equation that determins the values of $k$ and $\gamma$.
From Eq.~\eqref{constant_fixation} we obtain that
\begin{equation}
 4(\varepsilon_\mathbf{p}^-)^2 + 4\varepsilon_\mathbf{p}^- \cfrac{B_\mathbf{p}}{A_\mathbf{p}} = -(1+k^2)\gamma^2,
\end{equation}
or, equivalently,
\begin{equation}
 B_\mathbf{p} = -\left[\varepsilon_\mathbf{p}^- + \cfrac{(1+k^2)\gamma^2}{4\varepsilon_\mathbf{p}^-}\right]A_\mathbf{p}.\label{ba_connection}
\end{equation}
Still, we a way to determine $A_\mathbf{p}$.
In order to obtain it, let us consider $\det_s{\check{G}(\tau,\tau)}$.
On one hand, using the decomposition~\eqref{matrix_decomposition},
we can write
\begin{multline}
 \det_s{\check{G}(\tau,\tau)} = \cfrac{1}{4}\left[\left(\mathrm{tr}[\check{G}(\tau,\tau)]\right)^2
 \right. - \\ \* - \left. \vphantom{\left(\mathrm{tr}[\check{G}(\tau,\tau)]\right)^2}
 X_\mathbf{p}^2(\tau)-Y_\mathbf{p}^2(\tau)-Z_\mathbf{p}^2(\tau)\right],\label{det1}
\end{multline}
where
\begin{multline}
 \mathrm{tr}[\check{G}(\tau,\tau)] = -\cfrac{\mathrm{tr}[\check{U}_\mathbf{p}(\tau,0)\check{U}_\mathbf{p}(\beta,\tau)]}{2(\cosh{\beta\varepsilon_\mathbf{p}^+}+\cosh{\beta\kappa_\mathbf{p}})}
 = \\ \* =
 -\cfrac{\cosh{\beta\kappa_\mathbf{p}}}{\cosh{\beta\varepsilon_\mathbf{p}^+}+\cosh{\beta\kappa_\mathbf{p}}}.
\end{multline}
On the other hand,
\begin{multline}
 \det_s[\check{G}(\tau,\tau)] = \det_s\left[-\cfrac{\check{U}_\mathbf{p}(\tau,0)\check{U}_\mathbf{p}(\beta,\tau)}{2(\cosh{\beta\varepsilon_\mathbf{p}^+}+\cosh{\beta\kappa_\mathbf{p}})}\right]
 = \\ \* =
 \cfrac{\det_s{[\check{U}_\mathbf{p}(\tau,0)\check{U}_\mathbf{p}(\beta,\tau)]}}{4\left(\cosh{\beta\varepsilon_\mathbf{p}^+}+\cosh{\beta\kappa_\mathbf{p}}\right)^2} = \\ \* =
 \cfrac{1}{4\left(\cosh{\beta\varepsilon_\mathbf{p}^+}+\cosh{\beta\kappa_\mathbf{p}}\right)^2}.\label{det2}
\end{multline}
Combining~\eqref{det1} and ~\eqref{det2}, we arrive at
\begin{multline}
 X^2_\mathbf{p}(\tau) + Y^2_\mathbf{p}(\tau) + Z^2_\mathbf{p}(\tau) = \left(\cfrac{\sinh{\beta\kappa_\mathbf{p}}}{\cosh{\beta\varepsilon_\mathbf{p}^+} + \cosh{\beta\kappa_\mathbf{p}}}\right)^2
 = \\ \* =
 \cfrac{1}{4}\left(\tanh{\frac{\beta(\kappa_\mathbf{p}+\varepsilon_\mathbf{p}^+)}{2}} + \tanh{\frac{\beta(\kappa_\mathbf{p}-\varepsilon_\mathbf{p}^+)}{2}}\right)^2.\label{pseudospin_identity}
\end{multline}
Using Eqs.~\eqref{ansatz:y}, \eqref{ansatz:x}, \eqref{ansatz:z}, \eqref{ba_connection} and~\eqref{b_form}, we can express
\begin{multline}
 X^2_\mathbf{p}(\tau) + Y^2_\mathbf{p}(\tau) + Z^2_\mathbf{p}(\tau)
 = \\ \* =
 \left((\varepsilon_\mathbf{p}^+)^2 + \frac{\gamma^2(1+k)^2}{4}\right)\left((\varepsilon_\mathbf{p}^+)^2 + \frac{\gamma^2(1-k)^2}{4}\right)\cfrac{A_\mathbf{p}^2}{(\varepsilon_\mathbf{p}^+)^2}.\label{pseudospin_length}
\end{multline}
In this derivation, we also used the fact that the function $b(\tau) = \gamma k \times sn(\gamma\tau|k)$ satisfies additionally the equation
\begin{equation}
 \ddot{b}(\tau) + (1+k^2)\gamma^2 b^2(\tau) - b^4(\tau) - \gamma^4 k^2 = 0.
\end{equation}
Inserting Eq.~\eqref{pseudospin_length} into Eq.~\eqref{pseudospin_identity}, we can determine $A_\mathbf{p}$ up to a sign:
\begin{equation}
 A_\mathbf{p} = \pm \cfrac{|\varepsilon_\mathbf{p}^-|}{2}\cfrac{\tanh{\frac{\beta(\kappa_\mathbf{p}+\varepsilon_\mathbf{p}^+)}{2}} + \tanh{\frac{\beta(\kappa_\mathbf{p}-\varepsilon_\mathbf{p}^+)}{2}}}{\sqrt{\left((\varepsilon_\mathbf{p}^+)^2 + \frac{\gamma^2(1+k)^2}{4}\right)\left((\varepsilon_\mathbf{p}^+)^2 + \frac{\gamma^2(1-k)^2}{4}\right)}}
\end{equation}
Now, we can substitute $Y_\mathbf{p}(\tau) = A_\mathbf{p} b(\tau)$ into Eq.~\eqref{lgap:pseudospins} to obtain the self-consistency equation~\eqref{self_consistency}.
We shall notice, that the sign of $A_\mathbf{p}$ ($+$) can be uniquely fixed by requiring that, in the limit $k\rightarrow 1$, the self-consistency equation~\eqref{self_consistency} transforms into the self-consistency equation~\eqref{self_consistency:static} for the static case.

\subsection{Calculation of parameter $\kappa_\mathbf{p}$}

The final piece of the puzzle is to determine the value of parameter $\kappa_\mathbf{p}$ for the configuration $b(\tau)$ given by Eq.~\eqref{b_form}. In order to do that, we need to calculate $\mathrm{tr}{\left[\check{U}_\mathbf{p}(\beta,0)\right]}$.
It is convenient to make the unitary transformation of the basis
\begin{equation}
 \mathrm{tr}{\left[\check{U}_\mathbf{p}(\beta,0)\right]} = \mathrm{tr}{\left[\check{\BU}_0^\dagger \check{U}_\mathbf{p}(\beta,0) \check{\BU}_0\right]}
\end{equation}
with matrix
\begin{equation}
 \check{\BU}_0 = \cfrac{1}{2}
 \left(
    \begin{array}{cc}
     1 & i \\
     i & 1 
    \end{array}\right)
\end{equation}
In the rotated basis,
\begin{equation}
 \check{\BU}_0^\dagger \check\Sigma_1\check{\BU}_0 = \check\Sigma_1,\quad
 \check{\BU}_0^\dagger \check\Sigma_2\check{\BU}_0 = \check\Sigma_3,\quad
 \check{\BU}_0^\dagger \check\Sigma_3\check{\BU}_0 = -\check\Sigma_2,
\end{equation}
so that
\begin{multline}
 \check{\tilde{U}}_\mathbf{p}(\tau_2,\tau_1) = \check{\BU}_0^\dagger \check{U}_\mathbf{p}(\tau_2,\tau_1) \check{\BU}_0
 = \\ \* =
 \BT e^{\int_{\tau_1}^{\tau_2} d\tau \left(\varepsilon_\mathbf{p}^-\check\Sigma_2 + b(\tau) \check\Sigma_3\right)}.
\end{multline}
Consequently, the matrix $\check{\tilde{U}}_\mathbf{p}(\tau,0)$ satisfies the following differential equation:
\begin{equation}
 \cfrac{d}{d\tau} \check{\tilde{U}}_\mathbf{p}(\tau,0) = \left(\varepsilon_\mathbf{p}^-\check\Sigma_2 + b(\tau) \check\Sigma_3\right) \check{\tilde{U}}(\tau,0)\label{evolution:transformed}
\end{equation}
with the initial condition
\begin{equation}
 \check{\tilde{U}}_\mathbf{p}(0,0) = \check{\mathrm{I}}.\label{evolution:ic}
\end{equation}
Let us write the matrix $\check{\tilde{U}}_\mathbf{p}(\tau,0)$ explicitly as
\begin{equation}
 \check{\tilde{U}}(\tau,0) =
 \left(\begin{array}{cc}
            u_\mathbf{p}^{(1)}(\tau) & u_\mathbf{p}^{(2)}(\tau)\\
            v_\mathbf{p}^{(1)}(\tau) & v_\mathbf{p}^{(2)}(\tau)
       \end{array}\right)\label{evolution:explicit}
\end{equation}
Equation~\eqref{evolution:transformed} implies that each of the columns of $\check{\tilde{U}}(\tau,0)$ satisfies the system
\begin{subequations}\label{evolution:column1}
 \begin{align}
  \dot{u}_\mathbf{p}(\tau) & = b(\tau) u_\mathbf{p}(\tau) - i\varepsilon_\mathbf{p}^- v_\mathbf{p}(\tau),\label{evolution:u1}\\
  \dot{v}_\mathbf{p}(\tau) & = i\varepsilon_\mathbf{p}^- u_\mathbf{p}(\tau) - b(\tau) v_\mathbf{p}(\tau).\label{evolution:v1}
 \end{align}
\end{subequations}
We can substitute Eq.~\eqref{evolution:v2} into Eq.~\eqref{evolution:u1} or vice versa to obtain a closed equation for $u_\mathbf{p}$ or for $v_\mathbf{p}$ respectively:
\begin{subequations}\label{evolution:column2}
 \begin{align}
  \left[-\partial_\tau^2 + (\varepsilon_\mathbf{p}^-)^2 + b^2(\tau) + \dot{b}_y(\tau)\right]u_\mathbf{p}(\tau) = 0,\label{evolution:u2}\\
  \left[-\partial_\tau^2 + (\varepsilon_\mathbf{p}^-)^2 + b^2(\tau) - \dot{b}_y(\tau)\right]v_\mathbf{p}(\tau) = 0.\label{evolution:v2}
 \end{align}
\end{subequations}
Since we are interested in computing the trace, we only need to find $u_\mathbf{p}^{(1)}(\tau)$ and $v_\mathbf{p}^{(2)}(\tau)$.
The corresponding initial conditions can be determined by substituting Eqs.~\eqref{evolution:ic} and~\eqref{evolution:explicit} into the system~\eqref{evolution:column1}
\begin{subequations}\label{uv:ic}
 \begin{align}
  u_\mathbf{p}^{(1)}(0) = 1,&\qquad \dot{u}_\mathbf{p}^{(1)}(0) = 0,\label{u1:ic}\\
  v_\mathbf{p}^{(2)}(0) = 1,&\qquad \dot{v}_\mathbf{p}^{(2)}(0) = 0,\label{v2:ic}
 \end{align}
\end{subequations}
where we used the fact that $b(0) = 0$.

For the specific choice~\eqref{b_form} of $b(\tau)$ both equations of the system~\eqref{evolution:column2} are Lam\'e equations.
The general solution to Eq.~\eqref{evolution:u2} can be written as
\begin{multline}
 u_\mathbf{p}(\tau) = c_1 w_\mathbf{p}(\tau) \exp{\left[\int\limits_0^\tau d\tau^\prime \cfrac{\Omega_\mathbf{p}}{w_\mathbf{p}^2(\tau^\prime)}\right]}
 + \\ \* +
 c_2 w_\mathbf{p}(\tau) \exp{\left[-\int\limits_0^\tau d\tau^\prime \cfrac{\Omega_\mathbf{p}}{w_\mathbf{p}^2(\tau^\prime)}\right]},\label{u1:general}
\end{multline}
where
\begin{equation}
 \Omega_\mathbf{p} = |\varepsilon_\mathbf{p}^-|\sqrt{\left((\varepsilon_\mathbf{p}^+)^2 + \frac{\gamma^2(1+k)^2}{4}\right)\left((\varepsilon_\mathbf{p}^+)^2 + \frac{\gamma^2(1-k)^2}{4}\right)}
\end{equation}
and
\begin{equation}
 w_\mathbf{p}(\tau) = \sqrt{(\varepsilon_\mathbf{p}^-)^2 + \cfrac{1+k^2}{4} - \cfrac{b^2(\tau) + \dot{b}_y(\tau)}{2}}
\end{equation}
One can check that the initial conditions~\eqref{u1:ic} can be satisfied by the choice $c_1 = c_2 = (2w_\mathbf{p}(0))^{-1}$.

In regards to $v_\mathbf{p}^{(2)}$, we should point out that if one shifts $b(\tau) = \gamma k \sn{(\gamma\tau|k)}$ by the half-period $\omega = 2K(k)/\gamma$, $b(\tau + \omega) = -b(\tau)$. Thus,
\begin{equation}
 b^2(\tau + \omega) + \dot{b}_y(\tau+\omega) = b^2(\tau) - \dot{b}_y(\tau).
\end{equation}
As a result, the general solution to the equation~\eqref{evolution:v2} is also given by Eq.~\eqref{u1:general} but with $\tau+\omega$ substituted instead of $\tau$. Consequently, the initial conditions~\eqref{v2:ic} are satisfied by the choice $c_1=c_2=(2w_\mathbf{p}(\omega))^{-1}$.

If the integer number of periods $2\omega$ fits into the interval $[0, \beta]$, $w_\mathbf{p}(\tau + \beta) = w_\mathbf{p}(\tau)$.
As a result,
\begin{equation}
 u_\mathbf{p}^{(1)}(\beta) = v_\mathbf{p}^{(2)}(\beta) = \cosh{\left[\int\limits_0^\beta d\tau \cfrac{\Omega_\mathbf{p}}{w_\mathbf{p}^2(\tau)}\right]}.
\end{equation}
Comparing $\mathrm{tr}[\check{U}_\mathbf{p}(\beta,0)] = u_\mathbf{p}^{(1)}(\beta) + v_\mathbf{p}^{(2)}(\beta)$ with Eq.~\eqref{trace_form} we arrive at the conclusion that
\begin{equation}
 \kappa_\mathbf{p} = T \int\limits_0^\beta d\tau \cfrac{\Omega_\mathbf{p}}{w_\mathbf{p}^2(\tau)}.
\end{equation}
Finally, we can express the integral here in terms of complete elliptic integrals of the first and the third kind, thus arriving at Eqs.~\eqref{kappa} and~\eqref{n_ktilde}.

\end{document}